\documentclass[aps,prd,superscriptaddress,nofootinbib,tighten,preprint]{revtex4}
\pdfoutput=1
\usepackage[utf8]{inputenc}
\usepackage[english]{babel}
\usepackage{amsmath}
\usepackage{subfigure}
\usepackage{slashed}
\usepackage{amsfonts}
\usepackage{amssymb}
\usepackage{color}
\usepackage{cancel}
\usepackage{graphicx}
\newcommand{\be}{\begin{equation}}
\newcommand{\ee}{\end{equation}}
\newcommand{\bea}{\begin{eqnarray}}
\newcommand{\eea}{\end{eqnarray}}

\newcommand{\nn}{\nonumber}

\catcode`@=12

\begin{document}
\title{Common origin of baryon asymmetry, dark matter and neutrino mass}

\author{Anirban Biswas}
\email{anirban.biswas.sinp@gmail.com}
\affiliation{Department of Physics, Indian Institute of Technology Guwahati,
Assam 781039, India}
\affiliation{School of Physical Sciences, Indian Association for the Cultivation of
Science, 2A \& 2B Raja S.C. Mullick Road, Kolkata 700032, India}
\author{Sandhya Choubey}
\email{sandhya@hri.res.in}
\affiliation{Harish-Chandra Research Institute, HBNI, Chhatnag Road,
Jhunsi, Allahabad 211 019, India}
\author{Laura Covi}
\email{laura.covi@theorie.physik.uni-goettingen.de}
\affiliation{Institute for Theoretical Physics, Georg-August
University Göttingen, Friedrich-Hund-Platz
1, Göttingen, D-37077 Germany}
\author{Sarif Khan}
\email{sarifkhan@hri.res.in}
\affiliation{Harish-Chandra Research Institute, HBNI, Chhatnag Road,
Jhunsi, Allahabad 211 019, India}

\begin{abstract}
In this work, we explain three beyond standard model (BSM) phenomena, namely neutrino masses, the baryon asymmetry 
of the Universe and Dark Matter, within a single model and in each explanation the right handed (RH) neutrinos play the 
prime role. Indeed by just introducing two RH neutrinos we can generate the neutrino masses by the Type-I seesaw 
mechanism. The baryon asymmetry of the Universe can arise from thermal leptogenesis from the decay of lightest RH 
neutrino before the decoupling of the electroweak sphaleron transitions, which redistribute the $ B-L $ number into a 
baryon number. At the same time, the decay of the RH neutrino can produce the Dark Matter (DM) as an 
asymmetric Dark Matter component. The source of CP violation in the two sectors is exactly the same, 
related to the complex couplings of the neutrinos.
By determining the comoving number density for different values of the CP violation in the DM sector, we obtain 
a particular value of the DM mass after satisfying the relic density bound. We also give prediction for the 
DM direct detection (DD) in the near future by different ongoing DD experiments.    
\end{abstract}

\maketitle
\newpage

\section{Introduction}
\label{intro}

The Standard model (SM) is a concrete and successful theory which describes 
beautifully all the past and present particle physics measurements at colliders and 
at low energy experiments.
After the recent discovery of the Higgs boson, all the particles of the SM have been 
detected and are fully consistent so far with the SM predictions. 
In spite of this terrific success, we know though that we still do not have a complete
picture of particle physics.
Indeed on one side we have clear evidence for neutrino masses, by observing the neutrino
oscillation among the different flavours in atmospheric, solar and reactor neutrinos
\cite{Cowan:1992xc, Fukuda:1998mi, Ahmad:2002jz, Eguchi:2002dm, An:2015nua, 
RENO:2015ksa, Abe:2014bwa, Abe:2015awa, Salzgeber:2015gua, Adamson:2016tbq, 
Adamson:2016xxw}.
On the other hand, we know from astrophysical and cosmological data that in the Universe 
we need an additional matter component, apart for baryonic matter, to make up around 
$ 26\% $  of the total energy budget.
There are many evidences which support the existence of DM, mainly flatness of the 
galaxy rotation curves, the observation of bullet cluster and the CMB anisotropy
\cite{Sofue:2000jx, Clowe:2003tk, Harvey:2015hha, Bartelmann:1999yn, Hinshaw:2012aka, 
Ade:2015xua, Aghanim:2018eyx}.
In order to satisfy the observations DM has to be electrically neutral and stable or the decay 
life time has to be much longer than the age of the Universe (e.g. see ten point test of becoming the 
DM candidate in \cite{Taoso:2007qk}). No particle of the SM has the right
characteristics to provide the main Dark Matter component. Indeed neutrinos, which
are neutral and stable, are unfortunately too light and cannot be the sole Dark Matter
particle, see e.g. \cite{Choudhury:2018byy}.

Another big puzzle is that there exist an excess of baryonic matter over anti-baryonic
matter in the Universe and the baryon asymmetry has been measured very precisely by the 
satellite-based experiments WMAP and Planck \cite{Hinshaw:2012aka, Ade:2015xua, Aghanim:2018eyx}.
It turns out that the baryonic matter makes up approximately $5\% $ of the 
present energy density of the Universe and is approximately a factor $5$ less abundant 
than Dark Matter. In order to generate the baryon asymmetry of the Universe,
we need to satisfy the Sakharov conditions \cite{Sakharov:1967dj} including sufficient
C and CP violation and deviation from thermal equilibrium. These conditions are
difficult to realize within the SM and require generically new physics.

In this work, we mainly address the above mentioned three puzzles and try to solve them
in a unified manner through the presence of RH handed-neutrinos, which mediate with
the Dark Matter sector. 
Indeed the introduction of at least two RH neutrinos, singlets under the SM gauge group, 
allows to generate Majorana masses  for the light neutrinos by the Type-I seesaw 
mechanism \cite{Schechter:1980gr, Mohapatra:1979ia}. 
Moreover, it is well known that RH neutrinos can also generate the baryon asymmetry
of the Universe via leptogenesis \cite{Fukugita:1986hr, Covi:1996wh}. 
The lepton number asymmetry generated in the lightest RH neutrino decay can in fact be partially 
converted into a baryon asymmetry by sphaleron processes, which remain in thermal equilibrium 
until the electroweak phase transition~\cite{Kuzmin:1985mm}.

Regarding the Dark Matter, the most popular ways to generate the appropriate
density, like the freeze-out or freeze-in mechanisms, are usually independent from 
the neutrino sector and from baryogenesis.
But in our work, we will instead follow the paradigm of asymmetric Dark Matter (ADM)
also in order to explain the comparable densities of baryons and DM.
Indeed if the RH neutrinos not only have a Yukawa coupling with the light neutrinos 
and the SM Higgs, but also couple with the Dark Matter sector, they can decay also 
in the Dark Sector generating an asymmetry of a similar order. 
Many models of asymmetric Dark Matter have been proposed in the literature~
\cite{An:2009vq, Dutta:2010va, Falkowski:2011xh, Graesser:2011wi, McDermott:2011jp, 
Iminniyaz:2011yp, Kouvaris:2011fi, Arina:2011cu, Buckley:2011ye, Lin:2011gj, Blum:2012nf, 
Blennow:2012de, Okada:2012rm, Perez:2013nra, Petraki:2013wwa, Bhattacherjee:2013jca, 
Zurek:2013wia, Zhao:2014nsa, Bishara:2014gwa, Hamze:2014wca, Ibarra:2016fco, Narendra:2018vfw, 
Dong:2018aak, Ibe:2018juk, Dessert:2018khu, Ibe:2018tex},
but in this work we will give a new realization of the scenario and exploit and 
explore more in depth the connection to neutrino physics.

In order to solve the above mentioned problems in a common way, we extend the SM both in
the particle content as well as in the gauge group structure.  We add to the gauge group a
local $SU(2)_{D}$ interaction \footnote{WIMP type DM obeying SU(2) gauge
symmetry has been studied earlier in \cite{Banik:2015aya}.}
in the Dark sector and discrete $\mathbb{Z}_3$, $\mathbb{Z}_2$ groups,
ensuring the Dark Matter stability as well as forbidding the Majorana mass terms among the dark sector particles
and limiting the number of model parameters. 
The particle list is also enlarged to include two dark sector fermionic left handed doublets 
and their RH counterparts, singlet under the dark $SU(2)_{D}$, two scalar doublets of $SU(2)_D$ 
and two singlet RH neutrinos. We must introduce two $SU(2)_D$
fermionic doublets in order to cancel the Witten anomaly \cite{Witten:1982fp}. 
As we will see in the result section,
the RH neutrinos take part in generating the lepton asymmetry, Dark Matter asymmetry
and neutrino mass. One of the scalar doublets takes a vacuum expectation value, generating
a mass for the exotic fermions and also mixing with the SM Higgs. This opens up the
possibility to have DM scatterings mediated by the Higgs fields, which may be detected
in different ongoing and proposed direct detection experiments
\cite{Akerib:2015rjg, Aprile:2015uzo, Agnese:2014aze, lux2016, Aalbers:2016jon, Aprile:2017iyp, Cui:2017nnn}.
The other exotic scalar doublet does not obtain a non-zero vacuum expectation value and
plays a role similar to the inert doublet, participating in our case to the DM production.

The paper is organised in the following way. In section \ref{sec:model}, we describe all the details
of our model. The generation of the neutrino mass is discussed in section \ref{sec:neutrino-part}. 
In section \ref{sec:boltz}, we give the Boltzmann equations for both the dark sector and the leptonic 
sector, while section \ref{sec:result}, contains the full numerical result of the Boltzmann
equations. In section \ref{sec:DD}, the DM direct detection is addressed.  
Finally, in section \ref{sec:conclusion}, we conclude our work with an outlook on possible
signature at colliders.         

\section{model}
\label{sec:model}
In this article, we consider a hidden sector which has a local SU(2)$_D$ gauge
invariance. In this hidden sector we introduce four fermions $\psi_i$ ($i=1$ to 4)
whose left handed components transform like a doublet under SU(2)$_D$ while the right
handed counterparts are SU(2)$_D$ singlets. Therefore, we have a pair of SU(2)$_D$
doublets ${\Psi_{\alpha}}_L$ ($\alpha=1$, 2). In Table \ref{tab:tab2}, we show
the complete list of particles in the present model. Here we want to point that, since we
have an even number of fermionic SU(2)$_D$ doublets, our model is free from the
Witten anomaly \cite{Witten:1982fp}. In addition, we have two scalar doublets
in the hidden sector as well.
One of the scalar doublets $\eta_D$ does not get any vacuum expectation value (VEV)
while the remaining one ($\phi_D$) has a nonzero VEV and thus mixes with the SM
Higgs doublet $\phi_h$. Moreover, in order to have a stable DM candidate, we also impose
a discrete $\mathbb{Z}_3$ symmetry, and keep all the hidden sector fermions
as well as the inert doublet charged under $\mathbb{Z}_3$. These symmetries 
allow Majorana mass terms among the extra $SU(2)_D$ singlet fermions, which after the
breaking of the dark symmetry could switch on the conversion of the DM to anti-DM.
Therefore, to be on the safe side we introduce an additional $\mathbb{Z}_2$ symmetry to
forbid the Majorana mass terms among the extra fermions and reduce the possible couplings
of the second fermionic state.  Under $\mathbb{Z}_{2}$, $\psi_{2L}$, $\psi_{3R}$ and $\psi_{4R}$
are odd and the rest of the particles including the Standard Model particles are even.
Furthermore, we
have two right handed (RH) fermions ${N_i}$ ($i=$1, 2), singlets under both SM gauge group 
as well as SU(2)$_D$. These singlet fermions play the role of the RH neutrino and are
the only {\it connector} between the visible and the hidden sector, as long as the electroweak
and dark $SU(2)_D $ are unbroken and the mixing in the scalar sector vanishes.
In this sense our model is a special case in the class of neutrino(+Higgs) portal 
models~\cite{Falkowski:2009yz,Gonzalez-Macias:2016vxy,Escudero:2016tzx,Escudero:2016ksa,Batell:2017cmf,Chianese:2018dsz}. 
  
\subsection{Particle spectrum}
\def\I{i}
\begin{center}
\begin{table}[h!]
\begin{tabular}{||c|c|c||}
\hline
\hline
\begin{tabular}{c}
    Gauge\\
    Group\\ 
    \hline
    SU(3)$_{c}$\\ 
    \hline
    SU(2)$_{L}$\\ 
    \hline
    SU(2)$_D$\\
    \hline
    $\mathbb{Z}_{3} \times \mathbb{Z}_{2}$\\     
\end{tabular}
&
\begin{tabular}{c|c|c|c|c|c|c}
    \multicolumn{6}{c}{Fermion Fields}\\
    \hline
    ${\Psi_1}_{L} =(\psi_{1},\psi_{2})_L^{T}$ & ~$\psi_{1\,R}$~ & ~$\psi_{2\,R}$~ & 
${\Psi_2}_{L}=(\psi_{3},\psi_{4})_L^{T}$ & ~$\psi_{3\,R}$~ & ~$\psi_{4\,R}$~ & ~${N_i}$~ \\
    \hline
    $1$&$1$&$1$&$1$&$1$&1&1\\
    \hline
    $1$&$1$&$1$&$1$&$1$&1&1\\
\hline
    $2$&$1$&$1$&$2$&$1$&1&1\\
    \hline
    ($\omega,1$)&($\omega,1$)&($\omega,1$)&($\omega^2,-1$) & ($\omega^2,-1$)&($\omega^2,-1$)&($1,1$)\\
    
\end{tabular}
&
\begin{tabular}{c|c|c}
    \multicolumn{3}{c}{Scalar Fields}\\
    \hline
    $\phi_{h}$&$\phi_D$&$\eta_D$\\
    \hline
    ~~$1$~~&$1$&1\\
\hline
    ~~$2$~~&$1$&1\\
    \hline
    ~~$1$~~&$2$ & $2$\\
    \hline
    ~~($1,1$)~~&($1,1$) & ($\omega,1$)\\
     
\end{tabular}\\
\hline
\hline
\end{tabular}
\caption{List of hidden sector particles and {\it connector} particles
and their corresponding charges under various symmetry groups. All the particles
listed above have zero hypercharge except SM Higgs doublet $\phi_h$
which has hypercharge $Y = 1/2$.} 
\label{tab:tab2}
\end{table}
\end{center}  

\subsection{Lagrangian}

The SU(3)$_{c} \times {\rm SU(2)}_{L} \times {\rm SU(2)}_{D} \times {\rm U(1)}_{Y}
\times \mathbb{Z}_{3}$ invariant Lagrangian for our present model
is given by,
\begin{eqnarray}
\mathcal{L} & = & \mathcal{L}_{SM} 
+ i \,\overline{\Psi_k} \gamma^{\mu} D^k_{\mu}\,\Psi_k 
+ (D_{\mu}^D \phi_D)^{\dagger} (D^{D\mu} \phi_D) 
+ (D_{\mu}^D \eta_D)^{\dagger} (D^{D\mu} \eta_D) 
+ \left(y_{ij} \bar{L}_i \tilde{\phi_h} {N_j}_R + {\it h.c.}\right) \nn \\
&&
+ \left(\lambda_1 \overline{{\Psi_1}_{L}}\,\tilde{\phi_{D}}\,\psi_{1\,R} 
+ \lambda_2 \overline{{\Psi_1}_{L}}\,\phi_{D}\,\psi_{2\,R}
\right.\left. + \lambda_3 \overline{{\Psi_2}_{L}}\,\tilde{\phi_{D}}\,\psi_{3\,R} 
+ \lambda_4 \overline{{\Psi_2}_{L}}\,\phi_{D}\,\psi_{4\,R} + 
{\it h.c.}\right) \nn \\ &&
- \alpha_j \overline{{\Psi_1}_L} \eta_D {N_j}_R
+i\,\overline{{N_{j}}_R}\,\slashed{\partial}\,{N_{j}}_R- 
M_{j} \overline{{N_j}_R^c} {N_{j}}_R  
- \mathcal{V}(\phi_h,\phi_D,\eta_D) \,,
\label{Eq:lag}
\end{eqnarray}
where $i$ = 1 to 3 while $j$ runs from 1 to 2 and $D_{\mu}^D$ is the covariant
derivative for $SU(2)_{D}$. We assign the $\mathbb{Z}_3$ charges to the different fields
in such a way so that the lightest component of the dark doublet ${{\Psi_{1}}_L}$ becomes stable and
a viable DM candidate. Moreover, in the present work we are interested in $\psi_1$~\footnote{Like the SM 
lepton doublets, in dark sector too, we have assumed that between the components of a dark doublet the 
component with isospin $+1/2$ is the lightest one.} production from the decays of RH-neutrinos. 
Note that we can consider this state to have all real couplings by absorbing all the phases in $ \lambda_{1,2} $,
in the RH states $ \psi_{1,2 R} $, while the phases of $ \alpha_{1,2} $ can be absorbed into the doublets
$ \Psi_{1 L} $ and $ \eta_D $. Without loss of generality we can also consider the heavy Majorana masses
$ M_{1,2} $ to be real and positive, but redefining accordingly the heavy RH neutrino fields $ N_{1,2 R} $.

In the model there is also a second state, possibly $ \psi_3 $, which is stable due to the $ Z_2 $ and $ Z_3 $ 
symmetries and could contribute to the Dark Matter density.
To avoid both a substantial $ \psi_3 $ freeze-out density and a symmetric DM component from $ \psi_1 $,
the presence of the $ SU(2)_D$ gauge symmetry is crucial since it allows for an efficient annihilation
of the fermions, as long as they are not too heavy. We will discuss the precise value of the allowed
mass range later on. In this way, the main component of the DM in the Universe will be generated by 
the DM asymmetry in the RH neutrino's decays.   

Given the symmetries discussed above, the gauge invariant scalar potential has the following form
\begin{eqnarray}
\mathcal{V}(\phi_h,\phi_D,\eta_D) &=& -\mu_h^2 (\phi_h^{\dagger} \phi_h)
+ \lambda_h (\phi_h^{\dagger} \phi_h)^2 -\mu_D^2 (\phi_D^{\dagger} \phi_D)
+ \lambda_D (\phi_D^{\dagger} \phi_D)^2 \nn \\
&& + \mu_{\eta}^2 (\eta_D^{\dagger} \eta_D)
+ \lambda_{\eta} (\eta_D^{\dagger} \eta_D)^2
+ \lambda_{hD} (\phi_h^{\dagger} \phi_h)(\phi_D^{\dagger} \phi_D)
+ \lambda_{h\eta} (\phi_h^{\dagger} \phi_h)(\eta_D^{\dagger} \eta_D) \nn \\
&& + \lambda_{D1} (\phi_D^{\dagger} \phi_D)(\eta_D^{\dagger} \eta_D)
+ \lambda_{D2} (\phi_{D}^{\dagger} \eta_D) (\eta_D^{\dagger} \phi_{D})
+ \lambda_{D3} (\phi_{D} \eta_{D}^3 + h.c.)\,.
\label{Eq:vscalar}
\end{eqnarray}
Here, both the doublets $\phi_h$ and $\phi_{D}$ acquire VEVs and generate masses to 
SM particles and the hidden sector fermions 
after spontaneous breaking of SU(2)$_{L}\times {\rm U(1)}_{Y}$ and SU(2)$_{D}$ symmetries,
respectively.
In the unitary gauge, scalar doublets $\phi_h$ and $\phi_D$ take the following
form after symmetry breaking,
\begin{eqnarray}
\phi_h = \left(\begin{array}{c}
0 \\
\frac{v+h}{\sqrt{2}}  
\end{array}\right),\,\,\,
\phi_D = \left(\begin{array}{c}
0 \\
\frac{v_{D} + H}{\sqrt{2}}  
\end{array}\right) \,\,.
\label{doublet-vev}
\end{eqnarray}
In the scalar potential there is a mixing term between
the CP even neutral components of the two doublets $\phi_h$
and $\phi_D$, hence the gauge basis and mass eigenbasis will be different.
In $h$, $H$ basis the mass square mixing matrix will be as follows
\begin{eqnarray}
M^2_{scalar} = \left(\begin{array}{cc}
2 \lambda_{h} v^{2} & \lambda_{hD} v\, v_{D}\\
\lambda_{hD} v\, v_{D} & 2 \lambda_{D} v_{D}^{2}  
\end{array}\right),\,\,\,
\label{mass-square}
\end{eqnarray}
After diagonalising the above mass matrix we get
the physical masses and the corresponding physical
states which are linear combinations of gauge basis in
the following manner
\begin{eqnarray}
h_1 &=& h \cos \zeta - H \sin \zeta\,, \nn \\
h_2 &=& h \sin \zeta + H \cos \zeta\,,
\end{eqnarray}
where $\zeta $ is the mixing angle between $h_1$, $h_2$ and
the mixing angle can be expressed in terms of the Lagrangian
parameters in following way
\begin{eqnarray}
\tan 2\zeta = \frac{\lambda_{hD} v v_{D}}
{\lambda_{D} v_{D}^2 - \lambda_{h} v^2}\,.
\label{Eq:mixing_angle}
\end{eqnarray}
As mentioned above, after diagonalising the scalar mass matrix in Eq.\,(\ref{mass-square}),
we get the physical masses for the two neutral scalars as
\begin{eqnarray}
M_{h_{1}}^2 &=& \lambda_{h} v^2 + \lambda_{D} v_{D}^2 -
\sqrt{(\lambda_{D} v_{D}^2 - \lambda_{h} v^2)^2 + (\lambda_{hD}\, v\, v_{D})^2}\,, \nn \\
M_{h_{2}}^2 &=& \lambda_{h} v^2 + \lambda_{D} v_{D}^2 +
\sqrt{(\lambda_{D} v_{D}^2 - \lambda_{h} v^2)^2 + (\lambda_{hD}\, v\, v_{D})^2}\,.
\label{Eq:scalarmass}
\end{eqnarray}
We identify the lighter Higgs scalar as the SM-like Higgs observed at the LHC. 
Therefore, we take $ M_{h_1} = 126 $ GeV and consider small mixing angle $ \sin\zeta \leq 0.1 $
in order to ensure agreement with the Higgs signal strengths at the 
LHC~\cite{Khachatryan:2016vau, ATLAS:2017ovn, CMS:2018lkl}.

Now, we can express all the quartic couplings in terms of the physical
Higgs masses as,
\begin{eqnarray}
\lambda_{D} &=& \frac{M_{h_2}^2 + M_{h_1}^2 + (M_{h_2}^2 - M_{h_1}^2)\cos 2\zeta}
{4 v_{D}^2}\,, \nn \\
\lambda_{h} &=& \frac{M_{h_2}^2 + M_{h_1}^2 - (M_{h_2}^2 - M_{h_1}^2)\cos 2\zeta}
{4 v_{D}^2}\,, \nn \\
\lambda_{hD} &=& \frac{(M_{h_2}^2 - M_{h_1}^2)\sin 2\zeta}{2 v v_{D}}\,, \nn \\
\mu_h^{2} &=& \lambda_h {v^2} + \lambda_{hD} \frac{v_{D}^2}{2}\,, \nn \\
\mu_D^{2} &=& \lambda_D {v_D^2} + \lambda_{hD} \frac{v^2}{2}\,.
\end{eqnarray}
In the above expressions, all the quartic couplings have to be
within the perturbative regime which is $\lambda_i < $4\,$\pi$.

After the SU(2)$_D$ symmetry breaking, the DM candidate
$\psi_1 (= {\psi_1}_L \oplus {\psi_1}_R)$ will get mass which is
\begin{eqnarray}
M_{\psi_1} = M_{DM} = \frac{\lambda_{1} v_{D}}{\sqrt{2}}\,.
\label{dm-mass}
\end{eqnarray}
The other scalar doublet $\eta_D$, which has a nonzero $\mathbb{Z}_3$ charge,
will also get mass after breaking of both SM and hidden sector gauge symmetries.
In the present model, among the $\mathbb{Z}_3$ charged particles i.e.
hidden sector fermions and scalar $\eta_D$, we consider
the fermion $\psi_1$ as the lightest one. This is always possible
by tuning the couplings related to $\eta_D$ and $\lambda_{\alpha}$s ($\alpha=2$ to 4)
so that heavier $\mathbb{Z}_3$ charged particles decay to the lightest one
and thereby $\psi_1$ becomes stable. Thus, $\psi_1$ will be a viable DM candidate
in our model. Moreover, the invariance of the $\mathbb{Z}_3$ and $\mathbb{Z}_2$ symmetry actually
make the hidden sector fermion mass matrix diagonal, which means unlike the
quark or neutrino mixing in the SM, there is no mixing between the fermions
in different SU(2)$_D$ doublets.
%
%
%
\section{Neutrino mass}
\label{sec:neutrino-part}
In this work, as mentioned in the Model section (Section
\ref{sec:model}), instead of three right handed neutrinos we consider a minimal setting
and we add only two, which is sufficient to explain the current neutrino oscillation data. 
Therefore, in this framework, light neutrino masses are generated by the well known
Type-I seesaw mechanism, where the light neutrino mass matrix is related to the Dirac 
and Majorana mass matrices in the following way 
\begin{eqnarray}
m_{\nu} = - M_{D} M_{R}^{-1} M_{D}^{T} \,,
\label{neutrino-mass}
\end{eqnarray}
where $M_{D}$ is the Dirac mass matrix ($3 \times 2$)
while the Majorana mass matrix for the heavy right handed
neutrinos ${N_j}_R$ ($j=1,\,2$) is denoted by a $2 \times 2$
matrix $M_{R}$. In this work, for simplicity and without loss of
generality, we assume $M_R$ to be a diagonal matrix i.e. 
$M_R = {\rm diag}(M_{N_1}, M_{N_2})$.
One can also choose $M_{N_1}$, $M_{N_2}$ real and positive 
by redefining the phases of the spinors ${N_1}$ and ${N_{2}}$ in the 
mass eigenstate basis. Similarly, by redefining
the phases of the left handed neutrinos in the flavour basis,
we can remove the phases of one entire column of $M_D$ matrix.
So we consider all the elements of the first column of
the Dirac mass matrix ($M_{D}$) as real.
Therefore, in matrix form it looks like as follows,
\begin{eqnarray}
{M_{D}} =\frac{y_{i\,j}\, v}{\sqrt{2}} = \frac{v}{\sqrt{2}}
\left(\begin{array}{cc}
y_{ee} ~~&~~ y_{e\mu}^{R} - i y_{e\mu}^{I} \\
y_{\mu e}~~&~~ y_{\mu \mu}^{R} - i y_{\mu \mu}^{I}\\
y_{\tau e} ~~&~~ y_{\tau \mu}^{R} - i y_{\tau \mu}^{I} 
\end{array}\right) \,\,.
\label{dirac-mass-matrix}
\end{eqnarray}
We have computed the physical masses of light neutrinos
(eigenvalues) and mixing angles (eigenvectors) by diagonalising
the complex symmetric Majorana mass matrix $m_{\nu}$. All the elements
of $m_{\nu}$ are explicitly given in the Appendix \ref{app:mnu}.
In studying the neutrino phenomenology, we take into account
the observed values of neutrino oscillation parameters \cite{Capozzi:2016rtj}.
In this work, we mostly focus on the normal hierarchy
(NH) of the light neutrino masses, but a similar study
can be done for the inverted hierarchical scenario as well.
Bounds which we have considered to constrain the elements
of $m_{\nu}$ matrix are two mass square differences,
the three oscillation angles and the cosmological upper
bound on the sum of three light neutrino masses.
These bounds are as follows
\begin{itemize}
\item from the neutrino oscillation experiments there are tight
constraints on two mass square differences. For NH $3\sigma$ bounds
are as follows \cite{Capozzi:2016rtj},
\begin{eqnarray}
6.93 \leq \frac{\Delta m_{21}^2}{10^{-5}}
\,\, ({\rm eV^2})\,\, \leq 7.97\,\,
{\rm and} \,\,\, 2.37 \leq \frac{\Delta m_{31}^2}{10^{-3}}\,\,
({\rm eV^2})\,\,\leq 2.63\,\,.
\end{eqnarray}  
\item Three mixing angles namely, solar mixing angle
$\theta_{12}$, atmospheric mixing angle
$\theta_{23}$ and reactor mixing angle $\theta_{13}$
are also very well measured now. The
allowed values of mixing angles for NH in $3\sigma$ are
listed below \cite{Capozzi:2016rtj}
\begin{eqnarray}
30^{0} \leq &\theta_{12}& \leq 36.51^{0} \,\,,\nn \\
7.82^{0} \leq &\theta_{13}& \leq 9.02^{0} \,\,,\nn \\
37.99^{0} \leq &\theta_{23}& \leq 51.71^{0}\,, 
\end{eqnarray}
\item From the Planck data \citep{Ade:2015xua} there is a bound on the sum of 
light neutrinos masses which is $\sum_{i = 1,2,3}
m_{\nu_i} \leq 0.23$ eV.  
\end{itemize}
Moreover, we have not applied any bound on the CP violating phase
$\delta$, which is yet to be measured accurately by the different
ongoing and upcoming oscillation experiments like T2K \cite{Abe:2011sj},
T2HK \cite{Abe:2015zbg},
DUNE \cite{Acciarri:2015uup, Acciarri:2016ooe, Strait:2016mof, Acciarri:2016crz} and INO \cite{Kumar:2017sdq}. 
There is a hint of maximal CP violation
($\delta \sim - \frac{\pi}{2}$) from the T2K experiment \cite{Abe:2018wpn},
which excludes the CP conserving values of $\delta = 0$ or $\pi$, at 90\% C.L. However, 
there is some tension between the observed value of $\delta$ from T2K and NOvA \cite{NOvA:2018gge}.

In satisfying the above mentioned neutrino oscillation parameters,
we vary the model parameters, specifically, the elements of
the Dirac mass matrix (Eq.\,(\ref{dirac-mass-matrix}))
and the RH-neutrino masses in the following range, 
\begin{eqnarray}
10^{-4} &\leq y_{ee},\,y_{\mu e},
y_{\tau e},\,y_{e\mu}^{R} \leq & 10^{-2}\,\,, \nn \\
10^{-3} &\leq y_{\mu\mu}^{R},\,y_{\mu\mu}^{I},
\,y_{\tau\mu}^{R},\,y_{\tau\mu}^{I} \leq & 1\,\,, \nn\\
&& \hspace{-5.1cm} 10^{-6} \leq y_{e\mu}^{I} \leq 10^{-4}\,\,,\nn \\
10^{8}\, (10^{13}) &\leq M_{N_1}(M_{N_2})\,\,{\rm GeV}\,\leq & 10^{11}\,(10^{15})\,\,.
\label{Eq:para_ranges}
\end{eqnarray}
In Eq.\,\,(\ref{dirac-mass-matrix}), all the elements in the first column
are real and positive (can be made by phase rotation) while the elements in the second column
can have both positive as well as negative values. However,
to make the CP asymmetry parameter positive we consider
only positive values of all the elements in the second
column of the Dirac mass matrix, i.e. in the above
$y^{R,I}_{\alpha \mu} = |y^{R,I}_{\alpha \mu}|$, ($\alpha=e,\,\mu,\,\tau$).
Here, we take hierarchical RH-neutrino masses,
which is evident from the ranges of $M_{N_1}$ 
and $M_{N_2}$ that we have chosen in Eq.\,\,(\ref{Eq:para_ranges}).
The hierarchical scenario for the RH-neutrino masses implies
that we have to consider the lepton as well as DM asymmetry
generated from the decay of lightest RH-neutrino (${N_1}$) only.
Indeed the asymmetry generated from the decay of ${N_2}$ will be
washed out by the decay as well as inverse decay of ${N_1}$
which is still in thermal equilibrium during $T\sim M_{N_2}$. 
In the result section (Section \ref{sec:result}), 
we will see that there exist correlations among the parameters of the
neutrino sector and hence not the entire adopted ranges in
Eq.\,\,(\ref{Eq:para_ranges}) are allowed by the neutrino oscillation
data and the requirement of successful leptogenesis.

\section{Boltzmann Equation for studying lepton asymmetry
and Dark Matter asymmetry}
\label{sec:boltz}

As we have already mentioned earlier, in this work our goal is to generate
an asymmetry in the dark sector following the idea of leptogenesis in
the visible sector. In other words, the asymmetries in both sectors may have
a common source i.e.\,\,they can be generated from the CP-violating out-of-equilibrium
decay of the lightest RH-neutrino ${N_1}$.
Therefore, we need to solve (at least) three Boltzmann equations simultaneously: 
one gives the number density for the lightest RH-neutrino $N_1$ and the other two
will govern the asymmetries of the visible and dark sectors.
As mentioned in \cite{Falkowski:2011xh}, depending on the decay width and the
mass of the RH neutrinos, the source of dark and visible sector asymmetry
production can be divided in to two regimes. 
In the case  $\Gamma_{N_1} \ll M_{N_1}$, we are in the narrow width approximation
and the RH neutrino couples very weakly to the thermal bath, so that it is strongly 
out of equilibrium. On the other hand, for  $\Gamma_{N_1} \simeq M_{N_1}$,
we are in  the large washout or transfer regime where the wash-out processes
mediated by $ N_1$ are not negligible.
In the current work, we mostly focus on the RH-neutrino mass around
$10^{9}$ GeV and from the light neutrino mass constraint which is $\sim 10^{-11}$
GeV we have very small Yukawa couplings in the Dirac mass, hence we are 
in the narrow width approximation regime with  $\Gamma_{N_1} << M_{N_1}$.
Moreover, for this assumption we neglect the transfer diagrams~\cite{Falkowski:2011xh}. 

The relevant Boltzmann equations which we need to solve
for the narrow width approximation regime are as follows,
\begin{eqnarray}
\frac{d Y_{N_1}}{d z} &=& - \frac{z}{s H(M_1)} \left[ \left(\frac{Y_{N_1}}{Y_{N_1}^{eq}} - 1\right)
\times \left( \gamma_{D_1} + 2 \gamma^1_{\phi,s} + 4 \gamma^1_{\phi,t}  \right) \right]\,\,,
\label{Eq:BE_yn}\\
\frac{d Y_{\Delta l}}{d z} &=& - \frac{\Gamma_{N_1}}{H(M_1)} \left[ \epsilon_{l}
\frac{z K_{1}(z)}{K_{2}(z)} \left(Y_{N_1}^{eq} - Y_{N_1}\right) +  {{\rm Br}_l}
\frac{z^{3} K_{1}(z)}{4} Y_{\Delta l} \right]\,\,,
\label{Eq:BE_yl}\\
\frac{d Y_{D}}{d z} &=& - \frac{\Gamma_{N_1}}{H(M_1)} \left[ \epsilon_D
\frac{z K_{1}(z)}{K_{2}(z)} \left(Y_{N_1}^{eq} - Y_{N_1}\right) +  {{\rm Br}_D}
\frac{z^{3} K_{1}(z)}{4} Y_{D} \right]\,\,,
\label{Eq:BE_yd}
\end{eqnarray}
where the first equation represents the evolution of the comoving yield
$Y_{N_1} $ of ${N_1}$.
The yield of a species is defined as the actual number density of that species
divided by the entropy density of the Universe. If there is no interaction then
the yield of a species remains unaltered, as the expansion of the Universe
dilutes the number density and the entropy density in the same way.

The R.H.S. of the Boltzmann equation for ${N_1}$ describes the possible ways 
to change the number density of ${N_1}$. The quantity ${\gamma_D}_1$ is related to the total 
decay width of ${N_1}$ i.e. the decay of ${N_1}$ into both visible and dark sectors. 
On the other hand, $\gamma^1_{\phi,s}$ and $\gamma^1_{\phi,t}$ are related to
$s$-channel and $t$-channel scattering of ${N_1}$ mediated by $\phi_h$, which
can also lead to the destruction or production of ${N_1} $.
The expressions of $\gamma_{D_1}$, $\gamma^1_{\phi_h,s}$ and $\gamma^1_{\phi_h,t}$
are given in the Appendix \ref{app:b}. The second and third equations are
the evolution equations for the lepton asymmetry and Dark Matter asymmetry, respectively.
The first term in the R.H.S. of Eq.\,\,(\ref{Eq:BE_yl}) (Eq.\,\,(\ref{Eq:BE_yd})))
is the source term of lepton (Dark Matter) asymmetry from $N_1$ decay, while
the second term represents the washout effects on the created lepton (Dark Matter)
asymmetry due to the inverse decays of $N_1$. 
In Eqs.\,\,(\ref{Eq:BE_yl}, \ref{Eq:BE_yd}), ${{\rm Br}_{l}}$ and ${{\rm Br}_D}$
are the branching ratios of RH neutrino $N_1$ decay to leptonic sector
and dark sector, respectively. 
In the above equations the CP asymmetry parameters $\epsilon_{l,\, D}$ are zero 
at tree level. However, by considering both tree level and one loop level diagrams 
(vertex correction and wave function correction, see Fig.\,2 of \cite{Iso:2010mv}), 
non-zero values for the CP-asymmetry 
parameters $ \epsilon_{l,\, D} $ arise due to the interference between tree level 
and one loop level diagrams. The CP asymmetry
parameter for the visible sector is defined as \cite{Covi:1996wh, Falkowski:2011xh} 
\begin{eqnarray}
\epsilon_{l} &=& \frac{\Gamma({N_1} \rightarrow L \phi_{h}) -
\Gamma({N_1} \rightarrow \bar{L} \phi^{\dagger}_{h})}{\Gamma_{N_1}}\,\,, \nn \\
&=& \frac{M_{N_1}}{16 \pi M_{N_2}}\, \frac{{\rm Im}
\left[3 \left((y^\dagger y)^{\star}_{12}\right)^{2} + 2 \alpha_{1}^{\star}
\alpha_2 (y^\dagger y)^{\star}_{12} \right]}{\left[(y^{\dagger}y)_{11}
+ \alpha_{1} \alpha_{1}^{\star} \right]}\, ,
\label{epsilon-l}
\end{eqnarray}
where we have normalized to the total RH neutrino decay rate and summed over the  
lepton flavours, i.e. $\epsilon_{l} = \sum_{\alpha} \epsilon_{l}^{\alpha}$.
In Eq.~(\ref{epsilon-l}) we include contributions from the vertex and wave-function diagrams with 
virtual SM states as in classic leptogenesis, see e.g. \cite{Covi:1996wh}, and also the 
contribution
from the wave-function diagram with virtual dark sector states.

Similarly, the CP asymmetry in dark sector is defined as \cite{Falkowski:2011xh}
\begin{eqnarray}
\epsilon_D &=& \frac{\Gamma({N_1} \rightarrow {\psi_1}_{L}\,\eta_{D}) -
\Gamma({N_1} \rightarrow \overline{{\psi_1}_{L}}\,\eta_D^{\dagger})}{\Gamma_{N_1}}\,\,, \nn \\
&=& \frac{M_{N_1}}{16 \pi M_{N_2}}\,
\frac{{\rm Im}\left[2 \alpha_{1}^{\star} \alpha_2 (y^\dagger y)^{\star}_{12} 
+ 3 (\alpha_{1}^{\star} \alpha_2)^{2} \right]}{\left[(y^{\dagger}y)_{11} + 
\alpha_{1} \alpha_{1}^{\star} \right]}\,. 
\label{epsilon-chi}
\end{eqnarray}
In the case of Dark Matter, we have in an analogous way included contributions from 
the vertex and wave-function
diagrams from the dark sector and only the wave-function diagram from the leptons.
The total decay width of the RH-neutrino $N_{1}$ is given by
\begin{eqnarray}
\Gamma_{N_1} = \frac{M_{N_1}}{8 \, \pi}
\left[(y^{\dagger} y)_{11} + |\alpha_1|^2 \right]\,.
\label{Eq:gamma_tot_N1} 
\end{eqnarray}
From Eqs.\,\,(\ref{epsilon-l}, \ref{epsilon-chi}) we see that both $\epsilon_{l}$ and 
$\epsilon_D$ are determined by the Yukawa couplings $y_{ij}$ and RH-neutrino masses.
One very important thing to stress again is that in the dark sector we can absorb the phases 
of the couplings $\alpha_j$ ($j=1$, 2) by redefining the phases of ${\psi_1}_{L}$ and 
the complex scalar doublet $\eta_D$. Indeed, as in the case of a single generation
of fermions, no CP phase is physical in the DM sector.
Therefore the only source of CP violation for both the leptonic and DM sectors 
are the imaginary parts of the lepton-neutrino Yukawa couplings.
We have then
\begin{eqnarray}
\frac{\epsilon_l}{\epsilon_D} &=& 
\frac{{\rm Im}\left[3 \left((y^\dagger y)^{\star}_{12}\right)^{2} + 2\, \alpha_{1} \alpha_2 (y^\dagger y)^{\star}_{12} \right]}{
2\, \alpha_{1} \alpha_2\, {\rm Im}\left[(y^\dagger y)^{\star}_{12} \right]}
= 1 + \frac{{\rm Im}\left[3 \left((y^\dagger y)^{\star}_{12}\right)^{2} \right]}{
2\, \alpha_{1} \alpha_2 \, {\rm Im}\left[(y^\dagger y)^{\star}_{12} \right]}
= 1 +  3 \frac{\sum_\beta y_{\beta e} y^R_{\beta \mu}}{\alpha_{1} \alpha_2}
 \,. 
\label{epsilon-ratio}
\end{eqnarray}
We see that in the particular case when $ (y^\dagger y)^{\star}_{12} $ is purely imaginary,
i.e. $ y^R = 0 $, 
or generically when $ \alpha_1 \alpha_2 \gg \sum_\beta y_{\beta e} y^R_{\beta \mu} $, 
the two CP violation parameters are equal and we can expect a similar asymmetry in the 
two sectors, as long as the wash-out processes are negligible.
This is indeed not difficult to achieve as the Yukawas connected to the first generation
of the SM $ y_{\beta e} $ have to be small to give the small mixing angle $ \theta_{13} $.
From the matrix in Eq.~(\ref{dirac-mass-matrix}), we have that
\begin{equation}
 (y^\dagger y)^{\star}_{12} = y_{ee} ( y_{e\mu}^{R} + i y_{e\mu}^{I}) +
 y_{\mu e} ( y_{\mu\mu}^{R} + i y_{\mu\mu}^{I}) + y_{\tau e} ( y_{\tau\mu}^{R} + i y_{\tau\mu}^{I}) 
\end{equation}
so that this quantity is purely imaginary when the second column of the Dirac mass
matrix is purely imaginary and only six real Yukawa parameters remain. 
Note that in this limit, the CP  violation in the RH neutrino decay can still be large,
while the light neutrino mass matrix is real and the Dirac phase is therefore vanishing.
We have checked that with only imaginary components in the second column of the
Dirac mass matrix (see Eq.\,(\ref{dirac-mass-matrix})) and also in the same range
of the parameters value as given in Eq.\,(\ref{Eq:para_ranges}), one can easily obtain the three neutrino 
mixing angles and the  two mass square differences
in their observed $3\sigma$ ranges. For this particular choice of parameters,
one can easily estimate the value of the lepton CP asymmetry parameter ($\epsilon_l$).
In this case $\epsilon_l$ takes the following form,
\begin{eqnarray}
\epsilon_l = \frac{M_{N_1}}{ 8 \pi M_{N_2}} \frac{\alpha_2}{\alpha_1} {\rm Im}\left[ (y^{\dagger} y)_{12}^{\star}  \right]
\end{eqnarray}
for large $ \alpha_2/\alpha_1 $  and $ (y^\dagger y)_{11} < \alpha_1^2 $.
If we take ${\rm Im} (y^{\dagger} y)^{\star}_{12} \sim 2 \times 10^{-3}$,
$M_{N_1}/M_{N_2} \sim 10^{-4}$ and $\alpha_2/\alpha_1 \sim 10^2$, then we obtain
$\epsilon_l \sim 10^{-6}$ sufficient to generate the lepton asymmetry of the Universe,
as we will show later. One important conclusion we can draw for this scenario is that although 
there is no low scale CP violation~\footnote{If the light neutrino mass 
matrix is real, the Dirac phase $\delta $ is vanishing, but the Majorana phases can still be non-trivial
if the eigenvalues of the mass matrix have different sign. This indeed happens if one column of
the Yukawa matrix is imaginary, see the discussion in Appendix A.}, there still exist
a sufficiently large high scale CP violation by which we can generate lepton 
asymmetry of the Universe and henceforth the observed baryon asymmetry of
the Universe.

Another limiting case is when the real and imaginary parts $ y^{R,I}_{i \mu} $ in the
second column of the Yukawa matrix are equal and large. In that case it is  
$ ((y^\dagger y)_{12}^\star )^2 $ which becomes purely imaginary and can even
dominate the CP violation in the leptonic sector. In that case the dark CP violation
parameter $ \epsilon_D $ can be substantially smaller, but we can compensate
the smaller number density of the Dark Matter by increasing its mass and still
satisfy the CMB constraint.
Moreover, note that we can take large value of $\alpha_{2}$ ($\sim\mathcal{O}(1)$) without 
violating any constraint, but a large value of $\alpha_{1}$ will violate the narrow width 
approximation ($\Gamma_{N_1}\ll M_{N_1}$) via Eq.\,\,(\ref{Eq:gamma_tot_N1}).
Therefore, remaining within the narrow width approximation, but at the same time
aiming to increase the CP asymmetry parameter $\epsilon_l$
for the production of sufficient lepton asymmetry, we vary
the two parameters $\alpha_1$ and $\alpha_2$ in the following ranges:
\begin{eqnarray}
10^{-3} &\leq \alpha_1 \leq & 10^{-2}\,\,, \nn \\
0.3 &\leq \alpha_2 \leq & 1.0\,\,.
\label{al_1-al_2}
\end{eqnarray}

By solving the above three Boltzmann equations, we obtain the lepton and DM asymmetries 
as a function of the temperature of the Universe. For leptogenesis, we need to calculate
the lepton asymmetry before the sphaleron decoupling temperature
$T_{sph}\sim150$ GeV \cite{Plumacher:1996kc, Iso:2010mv} because the produced lepton
asymmetry has to be converted into the observed baryon asymmetry via sphaleron transitions 
in equilibrium.  The conversion
factor from $Y_{\Delta l}$ to $Y_{B}$ is given by \cite{Khlebnikov:1988sr}
\begin{eqnarray}
Y_{B} = \frac{8 N_{f} + 4 N_{\phi_h}}{22 N_{f} + 13 N_{\phi_h}} Y_{\Delta l}\,,
\end{eqnarray} 
where $N_f$ is number of generations of quarks and leptons
while $N_{\phi_h}$ being number of scalar doublets in the visible sector. 

Regarding the Dark Matter, we need to ensure that the symmetric DM component 
efficiently annihilates away leaving only the asymmetric component.
As the Dark Matter is charged under a Dark $SU(2) $ gauge group, this is not
difficult to achieve as long as the Dark Matter mass is not too large. Indeed
the dominant annihilation channel for an $SU(2) $ doublet is into the corresponding
gauge bosons, that subsequently annihilate/decay through the scalar sector into SM states
\footnote{In the same way, the symmetric component of $\psi_3$ annihilates away.}.
The generic expectation for the annihilation rate of a doublet is similar to that of
the Higgsino in supersymmetric models and it is enhanced by coannihilations~\cite{Edsjo:1997bg}
and even the Sommerfeld effect. In the case of Dark Matter mass of similar order to
the gauge boson mass and $ \lambda_2 \gtrsim \lambda_1 $, 
the coannihilation effect is dominant and we obtain~\cite{ArkaniHamed:2006mb}
\begin{equation}
\sigma_{\psi \bar\psi \rightarrow W_D W_D} \sim
\frac{ g_D^4}{128 \pi M_{DM}^2 } = 
2.5\times  10^{-9}\; \mbox{GeV}^{-2}  g_D^4 \left(\frac{M_{DM}}{1\;\mbox{TeV} }\right)^{-2}  \;
\end{equation}
which is slightly larger than the thermal cross section $ \langle \sigma v \rangle \sim 10^{-9} $ GeV$^{-2}$.
So for a Dark Gauge coupling of order one, the symmetric Dark Matter component
becomes important only at masses above $1$ TeV. Even heavier masses can be allowed
if the gauge bosons remain light and the Sommerfeld effect increases
the cross section  further~\cite{Baldes:2017gzw}.

For masses below the TeV range, where the annihilation is strong enough,
we can determine the Dark Matter relic density by computing 
Dark Matter asymmetry at the present epoch $T=T_0\sim \mathcal{O}(10^{-13})$ GeV ($z\rightarrow\infty$) 
and using the following relation \cite{Edsjo:1997bg},
\begin{eqnarray}
\Omega h^2 = 2.755 \times 10^{8} \left(\frac{M_{DM}}{\rm GeV} \right) Y_{D} (z \rightarrow \infty)\,.
\label{relic-density-formula}
\end{eqnarray}
Earlier WMAP and now the Planck satellite have measured the DM relic density
very precisely and its present value at the $68\%$ C.L.\,\,is
\cite{Ade:2015xua}
\begin{eqnarray}
0.1172 \leq \Omega h^{2} \leq 0.1226\,.
\label{planck-limit}
\end{eqnarray}
Equating Eq.\,\,(\ref{relic-density-formula}) and Eq.\,\,(\ref{planck-limit})
one can find the mass range for the Dark Matter particle $M_{DM}$ which
reproduces the observed Dark Matter abundance.

Let us finally comment on the presence of the inert doublet $ \eta_D $, which is also produced
in the RH neutrino decay and, as it is heavier, could regenerate  a $ \psi_1 $ population after
freeze-out and after the $SU(2)_D, SU(2)_{EW} $ gauge symmetries  are broken, by decaying
 into a DM fermion and a lepton. The number density of $ \eta_D $ can be
 efficiently reduced by the cubic interaction with $ \phi_D $, which allows for
 semiannihilation~\cite{DEramo:2010keq} through the process $ \eta_D + \eta_D \rightarrow \eta_D^\star + h_{1,2} $. 
In this case semiannihilation is determined by the potential parameter $ \lambda_{D3} $,
which does not affect the stability of the vacuum and can be chosen large to allow for
a sufficiently large semiannihilation cross section.
 We assume here that due to such process only a negligible number density of the inert
 doublet is left after freeze-out.

\section{Results}
\label{sec:result}

\begin{figure}[h!]
\centering
\includegraphics[angle=0,height=9cm,width=12cm]{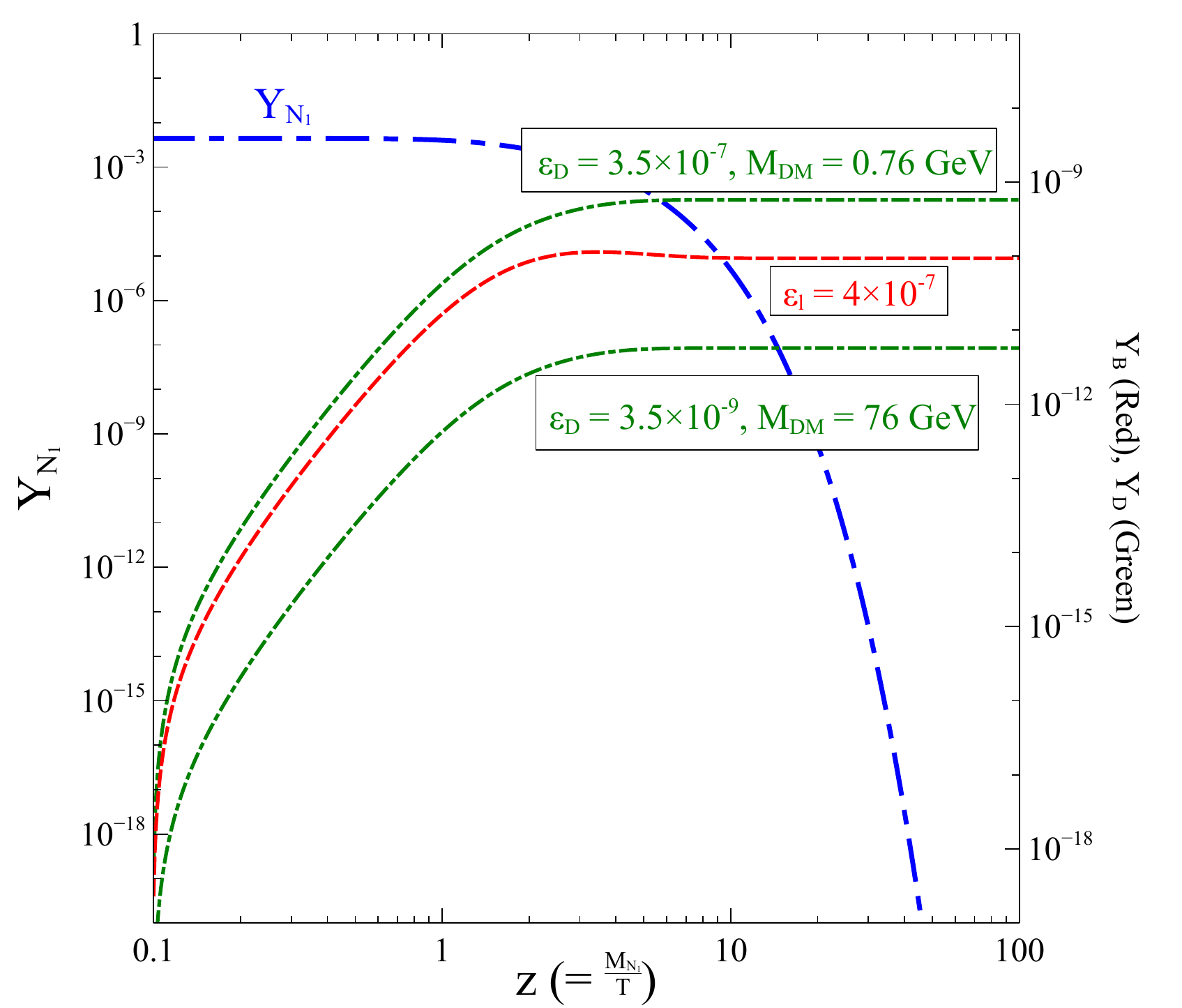}
\caption{Variation of baryon asymmetry of the Universe and the dark sector
asymmetry with z. In generating the figure we have assumed the following
values $M_{N_1} = 10^{10}$ GeV, $M_{N_2} = 10^{13}$ GeV,
$y^{\dagger} y = 10^{-6}$, $\Gamma_{N_1} \sim 398$ GeV,
$Br_{\chi} = 0.2$, $Br_{l} = 0.8$.}
\label{yn-yb-yx-z}
\end{figure}
In Fig.\,\ref{yn-yb-yx-z}, we show the production of the baryon asymmetry and the
Dark Matter asymmetry from the decay of RH neutrino $N_1$. When  the RH neutrino
starts decaying, both the lepton sector asymmetry and the dark sector asymmetry
grow fast to their final values. For $\epsilon_{l} = 4 \times 10^{-7}$
and taking the other parameters values as mentioned in the caption, we obtain
the correct value of the matter antimatter asymmetry of the Universe
which lies within the value measured by the Planck satellite.
For two values of $\epsilon_{D} = 3.5 \times 10^{-7}$ and $3.5 \times 10^{-9}$,
we can provide for a dark sector asymmetry which fulfils the total Dark Matter
abundance for a Dark Matter mass of 0.76 GeV and 76 GeV, respectively.
One interesting thing to note here is that for $\epsilon_{D} = 3.5 \times 10^{-7}$
(which is less than $\epsilon_l$) we are producing more
dark asymmetry than the lepton asymmetry. This is because we have considered
$Br_D < Br_{l}$ {\it i.e} washout effects for the dark sector are weaker than for the 
visible sector. So even for CP violation of the same order, we can obtain an
enhancement of the Dark Matter density compared to the baryon density
giving naturally $ \Omega_{DM} > \Omega_b $  for DM masses in the
range of the charm/bottom quark mass. 
In the subsequent figures we will see the dependence of the baryon and dark 
sector asymmetry on the model parameters and how they are related with 
neutrino oscillations. 

\begin{figure}[h!]
\centering
\includegraphics[angle=0,height=7cm,width=8cm]{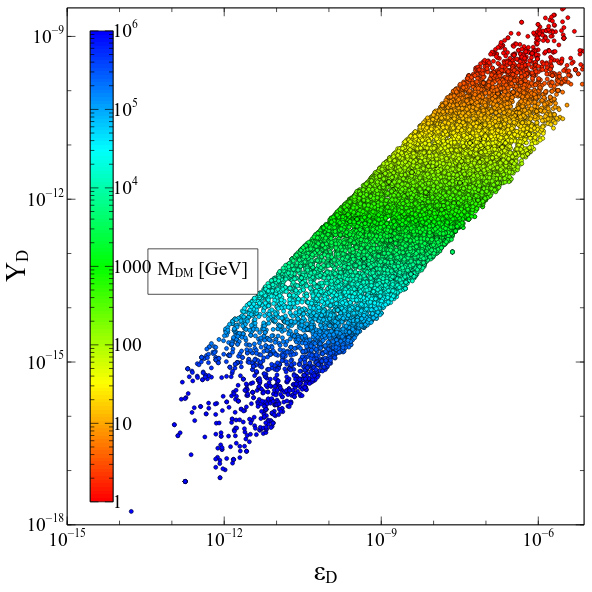}
\includegraphics[angle=0,height=7cm,width=8cm]{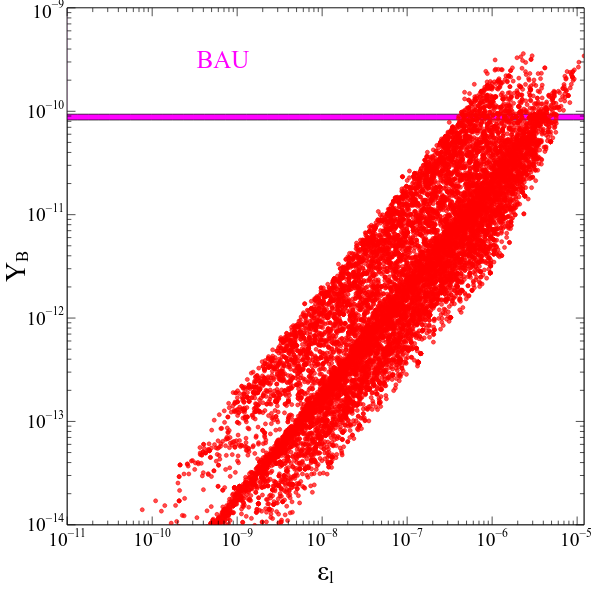}
\caption{Left panel (LP) shows the variation in the $Y_{D} - \epsilon_{D}$ 
plane and in the right panel (RP) variation in ($Y_{B}-\epsilon_{l}$) plane
has been shown. All the points satisfy neutrino oscillation data as mentioned in section
\ref{sec:neutrino-part}. All the parameters have been varied in the range as shown
in section \ref{sec:neutrino-part}.}
\label{scat-1}
\end{figure}
In the LP of Fig.\,\ref{scat-1}, we show the variation of the Dark Matter  yield with 
$\epsilon_{D}$. All the points satisfy the neutrino oscillation data.
The figure shows a sharp correlation between $\epsilon_{D}$ 
and $Y_{D}$, as expected in the narrow width regime.
For the particular value of the yield we determine the Dark Matter mass 
such that the DM energy density coincides with the observed value
using the expression as given in Eq.\,(\ref{relic-density-formula}).
We see that the present model can accommodate the right Dark Matter
abundance in the mass range $ M_{DM} = 1- 10^6 $ GeV, which is,
at least in the lower mass range, within the sensitivity of ongoing direct detection 
experiments. The large mass region above the TeV is disfavoured by 
the fact that the CP violation parameter has to be very suppressed 
and by the possible presence of a substantial symmetric DM component.

In the RP, we show the allowed region in the $Y_{B}-\epsilon_{l}$ plane. 
As expected, here also a direct correlation exist between these two quantities.
The narrow magenta band is the present day accepted value of the baryon anti-baryon
asymmetry of the Universe. The parameter values for which the baryon asymmetry
lies above the magenta line are ruled out, while below the line leptogenesis cannot
provide the full matter-antimatter asymmetry of the Universe.
We see that we need a CP violation parameter $ \epsilon_l \sim 10^{-6} $
in order to obtain the observed baryon asymmetry.
Moreover, comparing the LP with the RP, it is clear that we can indeed achieve the
right abundance of Dark Matter and baryons for the case $ \epsilon_l \sim \epsilon_D $
and then the Dark Matter mass is in the range $1-10$ GeV as expected.

\begin{figure}[h!]
\centering
\includegraphics[angle=0,height=7cm,width=8cm]{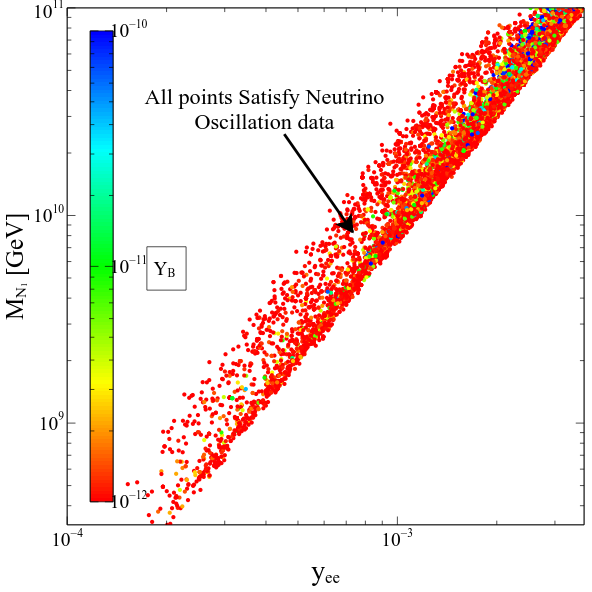}
\includegraphics[angle=0,height=7cm,width=8cm]{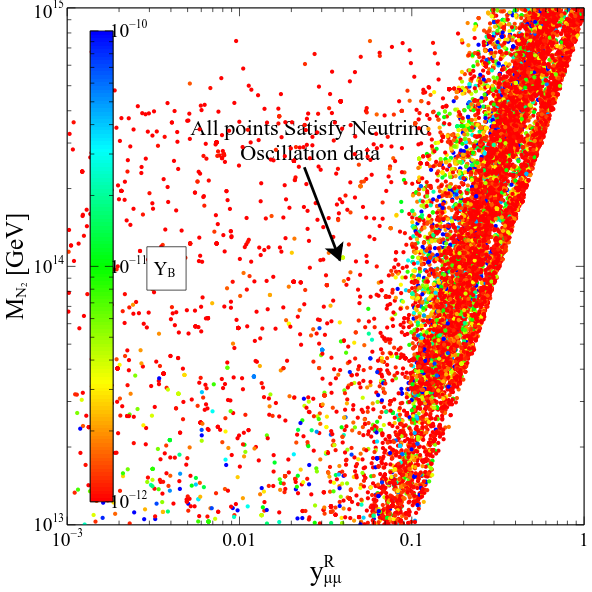}
\caption{LP (RP): Scatter plot in the $y_{ee}-M_{N_1}$ ($y_{\mu\mu}^{R}-M_{N_2}$) plane
after satisfying neutrino oscillation data as mentioned in section
\ref{sec:neutrino-part}. All the parameters have been varied in the range as shown
in section \ref{sec:neutrino-part}.}
\label{scat-2}
\end{figure}
In the LP and RP of Fig.\,\ref{scat-2}, we show the correlation among the
RH neutrino masses and the elements of the Dirac mass matrix, which is
required in order to satisfy the oscillation data.
We see that the ratio  $\frac{y_{ee}^2}{M_{N_1}}$ is fixed by 
neutrino oscillations, as it provides one of the mass scales in the light neutrino mass
matrix. Similar 
correlations of $M_{N_1}$ are present also for the other Yukawa parameters in the first column,
$ y_{\mu e}, y_{\tau e} $. Here we are generating the neutrino mass by the Type-I seesaw mechanism, 
hence, the elements in the first column of the Dirac mass matrix 
(see Eq.\,(\ref{dirac-mass-matrix}) and Appendix A)
are directly related to $M_{N_1}$. 
Similarly, the elements of the second column of the Dirac mass matrix are always 
suppressed by $1/M_{N_2}$ when they appear in the light neutrino mass matrix. 
Therefore, for the RP we also see a similar kind
of correlation with elements of the second column of Dirac mass matrix and $M_{N_2}$,
but in this case the correlation is less sharp.  Since the RH neutrinos $N_1$ and $N_2$ are hierarchical, a 
difference in the magnitude of the elements of the first and second column of the Yukawa 
matrix are arranged such that both the light neutrino mass squared differences we obtain are 
consistent with the neutrino oscillation data. 
In these plots, all points are allowed by neutrino oscillation data 
and some of them (indicated by blue colour) produce the observed
matter-antimatter asymmetry of the Universe and for rest we need
extra sources of baryogenesis. Indeed we have adjusted the imaginary part
of the elements of the second column of the Yukawa matrix such that both 
the lepton asymmetry as well as the light neutrino mass matrix are obtained,
consistent with their respective measured values.

\begin{figure}[h!]
\centering
\includegraphics[angle=0,height=7cm,width=8cm]{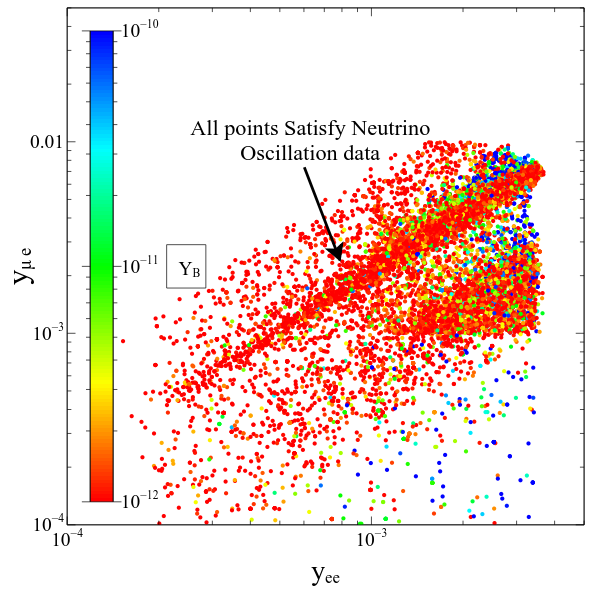}
\includegraphics[angle=0,height=7cm,width=8cm]{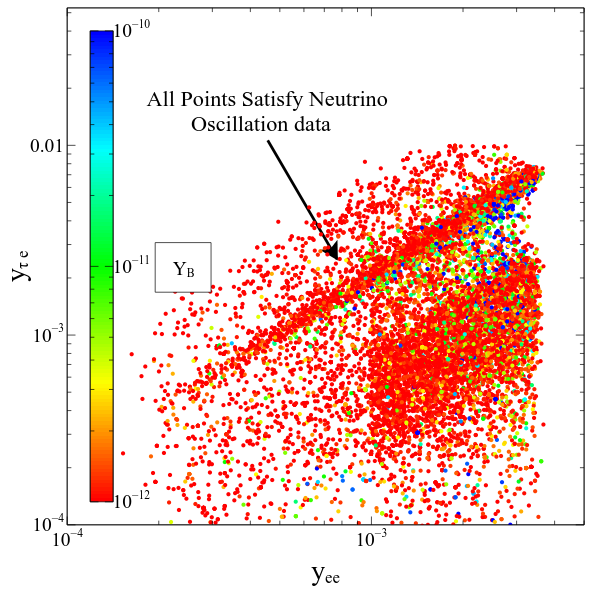}
\caption{LP (RP): Scatter plot in the $y_{ee}-y_{\mu e}$ ($y_{ee}-y_{\tau e}$) plane
after satisfying neutrino oscillation data as mentioned in section
\ref{sec:neutrino-part}. All the parameters have been varied in the range as shown
in section \ref{sec:neutrino-part}.}
\label{scat-3}
\end{figure}
In Fig.\,\ref{scat-3}, we show the allowed region after satisfying the neutrino
oscillation data in both LP and RP. A clear correlation exist among the parameters
in order to satisfy neutrino oscillation data.
Only the blue points are close to the current value of the Universe's baryon asymmetry,
while the red and green points give a too low value.

\begin{figure}[h!]
\centering
\includegraphics[angle=0,height=7cm,width=8cm]{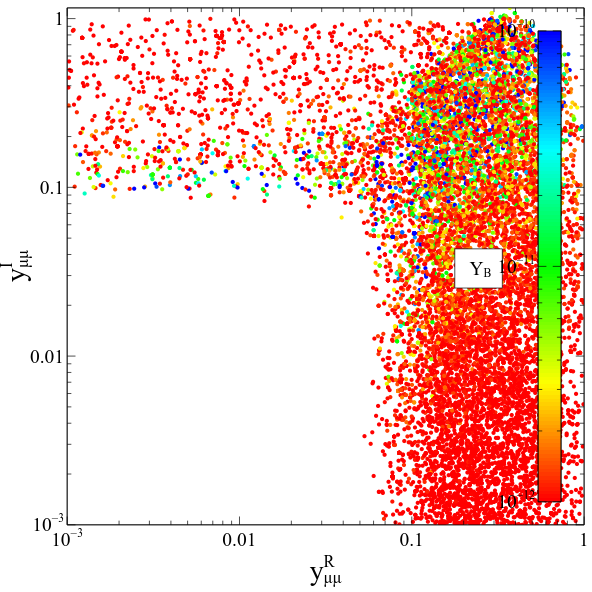}
\includegraphics[angle=0,height=7cm,width=8cm]{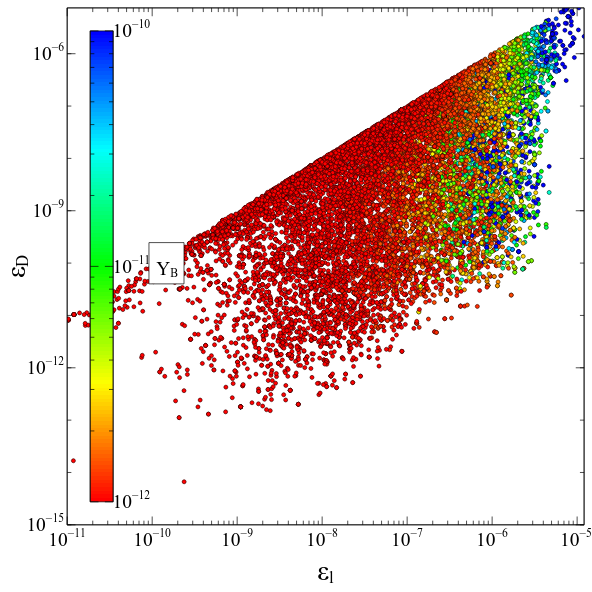}
\caption{LP (RP): Scatter plot in the $y_{\mu\mu}^{R}- y_{\mu\mu}^{I}$ 
($\epsilon_{l}-\epsilon_{D}$) plane
after satisfying neutrino oscillation data as mentioned in section
\ref{sec:neutrino-part}. All the parameters have been varied in the range as shown
in section \ref{sec:neutrino-part}.}
\label{scat-4}
\end{figure}

The LP of Fig.\,\ref{scat-4} shows explicitly the correlation between the real
and imaginary part of the same element of the Dirac mass matrix, i.e.
$y_{\mu\mu}^{R}$ and $y_{\mu\mu}^{I}$.
Either of them can give the right contribution to the light neutrino mass
matrix to fit the oscillation data, but only a substantial imaginary part
allows for non-vanishing CP violation and the production of a sufficiently
large lepton asymmetry.
This plot shows that the correct baryon asymmetry can be obtained
both if the real and imaginary part are equal and large or when the
real part is negligible and the imaginary part provides a substantial 
contribution to both the CP violation and the neutrino mass.
This two cases correspond to the limiting cases discussed earlier.

In RP of the same figure we show the allowed region in the $\epsilon_l - \epsilon_D$ 
plane. We see that our predicted baryon asymmetry
comes close to the measured value 
only for reasonably high values of $\epsilon_l$. Since there exists a sharp
correlation between $Y_{B}$ and $\epsilon_l$, lower $\epsilon_l$ results in 
production of lower lepton asymmetry.
As we discussed earlier, we expect $ \epsilon_D $ to be equal or less than
$ \epsilon_l $ and indeed this is also reproduced in this figure.
Note that while the baryon asymmetry is correctly given only in a quite
narrow region of the parameter space, a much wider range of $\epsilon_D$ 
is allowed as we can adjust the DM mass to match the Dark Matter abundance.
Generically the mass of the Dark Matter state is given by the VEV of the
Dark $SU(2)_D$ scalar doublet and can vary compared to the scale of
SM fermion masses. Nevertheless a natural range for the mechanism to work
are DM masses in the GeV to tens of GeV range, well below the WIMP 
mass scale around 1 TeV. 

\begin{figure}[t]
\centering
\includegraphics[angle=0,height=7cm,width=8cm]{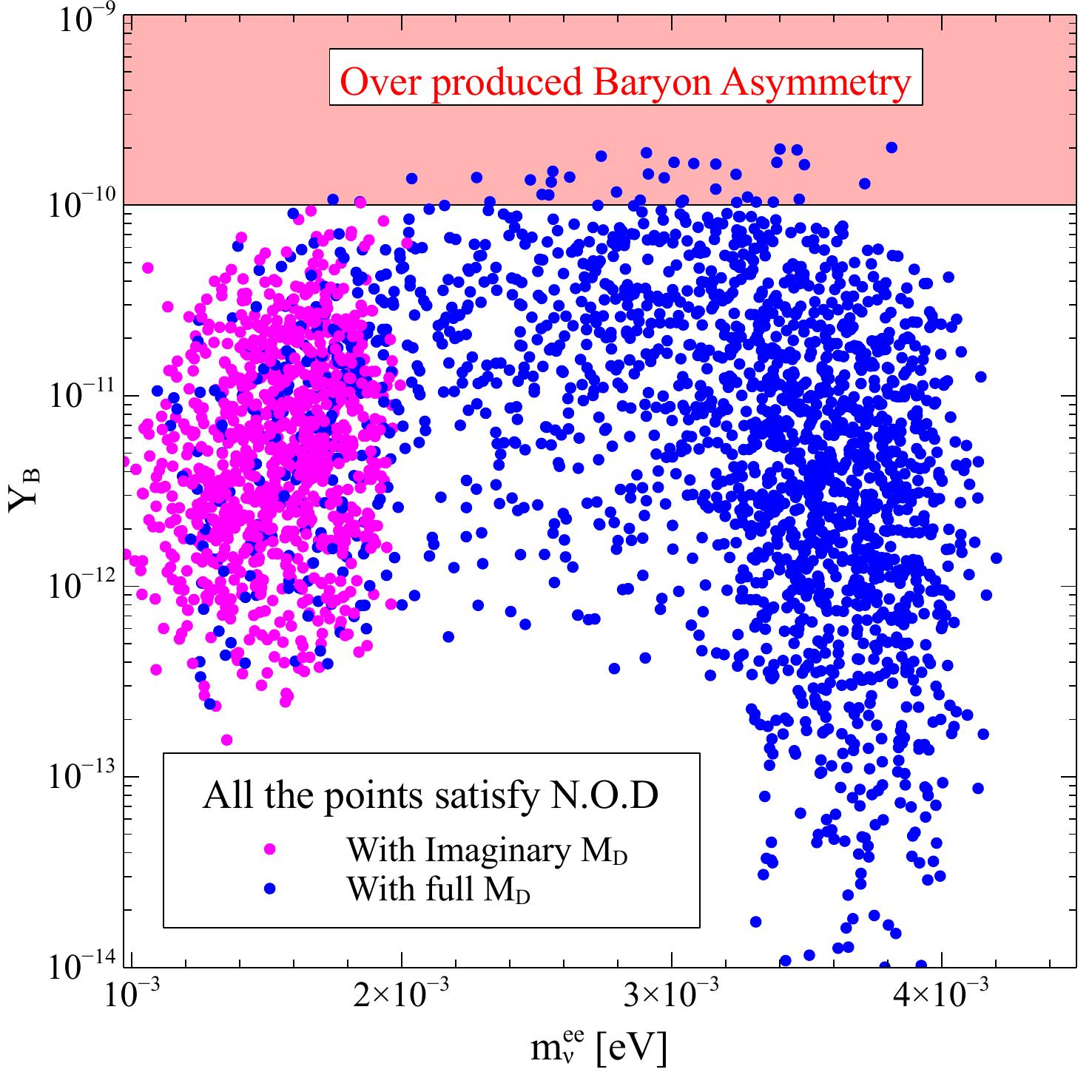}
\caption{Scatter plot in the $m_{\nu}^{ee}-Y_{B}$ plane
after satisfying neutrino oscillation data (N.O.D) as mentioned in section
\ref{sec:neutrino-part}. All the parameters have been varied in the range as shown
in section \ref{sec:neutrino-part}.}
\label{scat-5}
\end{figure}
In Fig.\,\,\ref{scat-5}, we give the allowed model points in the $m_{\nu}^{ee}-Y_{B}$
plane. Both the blue and magenta points satisfy fully the neutrino oscillation data and
the only difference between them is the choice of the Dirac mass matrix elements 
in Eq.\,\,\ref{dirac-mass-matrix}.
The blue points correspond to the general $M_{D}$ matrix with complex second column,
whereas the magenta points are obtained for a purely imaginary second column of
$M_{D}$ ({\it i.e.} $y_{i\mu}^{R} = 0$, $i = e,\,\, \mu,\,\, \tau$).    
Taking the Majorana phase angle convention for the light neutrinos as 
diag(1, $e^{i \alpha_{21}/2}$, $e^{i \alpha_{31}/2}$) and one zero mass eigenstate $ m_1=0$, 
we can write down the (1,1) element of the neutrino mass matrix ($m_{\nu}$), which
 generates the $0\nu \beta \beta$ decay, as
\begin{eqnarray}
m_{\nu}^{ee} = \left[s^4_{12} c^4_{13} m^2_2 + 2 m_2 m_3 s^2_{12} c^2_{13} s^2_{13}
\cos(\alpha_{31} - \alpha_{21} - 2 \delta_{CP}) + m^2_{3} s^4_{13}\right]^{1/2}
\end{eqnarray}
where $m_2,\,m_3$ are the light neutrino masses, $c_{ij} = \cos \theta_{ij}$, $s_{ij} = \sin \theta_{ij}$
are the mixing angles, while $\delta_{CP}$ is Dirac CP phase and $\alpha_{31}$, $\alpha_{21}$ are 
the Majorana phases~\footnote{Note that for one massless neutrino eigenstate, only one 
Majorana phase, given as the difference between $ \alpha_{31} $ and $ \alpha_{21} $, is physical}.  
Varying the mixing angles and the phases to cover $\cos(\alpha_{31} - \alpha_{21} - 2 \delta_{CP}) = \pm 1$,
we get a range of $m^{ee}_{\nu}$ for the normal hierarchy which lies in between $1$ meV and $4.3$ meV. 
This range is in complete agreement with the Fig.\,\,\ref{scat-5} which we obtain after satisfying the neutrino
oscillation data (as given in Section \ref{sec:neutrino-part}) at the time of diagonalising neutrino mass matrix 
$m_{\nu}$. 
Magenta points with a real light neutrino mass matrix correspond to $\delta_{CP} = 0$ and 
$(\alpha_{31}\,-\,\alpha_{21} ) = \pi$, so they point towards the lowest possible value of the 
neutrinoless double beta decay rate, while the small spread in the $Y_{B} - m^{ee}_{\nu}$ plane 
is mainly due to the variation of $\theta_{12}$ and $m_2$ within their experimental ranges. 
Hence, with a purely imaginary second column of $M_{D}$, we can satisfy both neutrino oscillation 
data and produce the baryon  and DM asymmetry of the Universe with the Dirac CP
phase $\delta_{CP} = 0 $ and  $\epsilon_{l} = \epsilon_{D}$.  This is the case of the minimal number 
of parameters in the neutrino sector.

In the case of a general $M_{D}$ matrix (shown by the blue points), we can obtain higher values of
$m_{\nu}^{ee}$ and hence it will be easier for $0\nu\beta\beta$ experiments to test this scenario.
The higher values of $m_{\nu}^{ee} \sim 4.2$ meV imply as well a Dirac CP phase
$\delta_{CP} = -\frac{\pi}{2}$ (in accordance with the recent data from the neutrino oscillation experiments) 
when $(\alpha_{31}- \alpha_{21})= \pi$. Therefore, with a generic $M_{D}$ matrix, we can satisfy 
the recent data of the neutrino oscillation parameters and produce the baryon and DM asymmetry of 
the Universe also with large Dirac CP phase, $\delta_{CP} \sim -\frac{\pi}{2}$.
 

\section{Direct Detection of Dark Matter}
\label{sec:DD}

\begin{figure}[h!]
\centering
\includegraphics[angle=0,height=5cm,width=8cm]{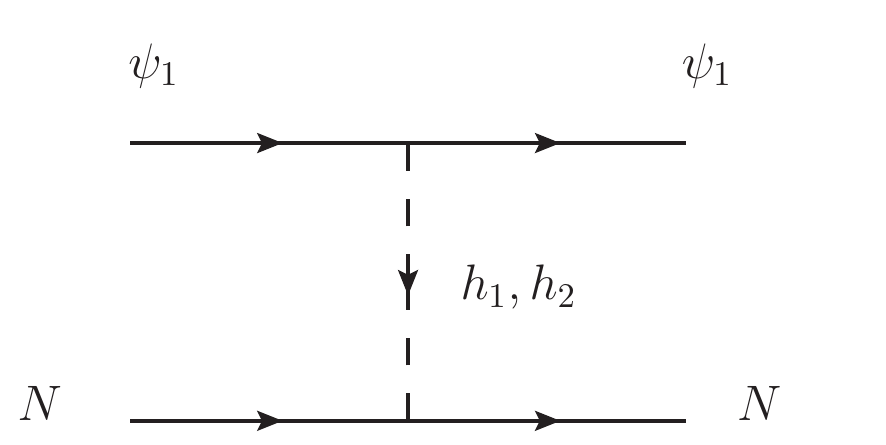}
\caption{Feynman diagram for the spin-independent
scattering cross section of Dark Matter with nucleon mediated
by both SM-like Higgs $h_1$ and hidden sector
Higgs $h_2$.}
\label{dd-fig}
\end{figure}

In our model, although DM only has a dark SU(2)$_D$ charge,
it can still {\it talk} to the visible sector through the exchange of
SM-like Higgs boson $h_1$ and dark sector Higgs $h_2$ respectively and the
corresponding Feynman diagram is shown in Fig.\,\ref{dd-fig}.

The expression for the spin-independent scattering cross section
between DM and nucleon mediated by scalars $h_1$, $h_2$ is
given by
\begin{eqnarray}
\sigma_{SI} = \frac{\mu_{red}^2}{\pi} \left[\frac{M_{N} f_{N}}{v}
\left(\frac{g_{\psi_1 \psi_1 h_2} \sin\zeta}{M_{h_2}^2}
+ \frac{g_{\psi_1 \psi_1 h_1} \cos\zeta}{M_{h_1}^2} \right) \right]^2\,,
\end{eqnarray}
where $\mu_{red} = \frac{M_{N} M_{DM}}{M_{N} + M_{DM}}$ is the
reduced mass and $f_N \sim 0.3$ \cite{Cline:2013gha}. The DM couplings with
the scalars have the following form
\begin{eqnarray}
g_{\psi_1 \psi_1 h_1} &=& - \frac{\lambda_1}{\sqrt{2}} \sin\zeta  = 
- \frac{m_{DM}}{v_D} \sin\zeta
\nn \\
g_{\psi_1 \psi_1 h_2} &=& \frac{\lambda_1}{\sqrt{2}}\cos\zeta = \frac{m_{DM}}{v_D} \cos\zeta\, ,
\end{eqnarray}
so we see that we have a negative interference and full cancellation
for equal masses of the dark and SM Higgs fields, where the mixing in the
scalar sector vanishes. Indeed we obtain
\begin{eqnarray}
\sigma_{SI} = \frac{\mu_{red}^2}{\pi} \left[\frac{M_{N} f_{N}}{v}
\frac{|\lambda_1| \sin 2 \zeta}{2 \sqrt{2} \; M_{h_1}^2}  \;
\left( 1 - \frac{M_{h_1}^2 }{M_{h_2}^2} \right)
 \right]^2\, .
\end{eqnarray}

\begin{figure}[t]
\centering
\includegraphics[angle=0,height=9cm,width=12cm]{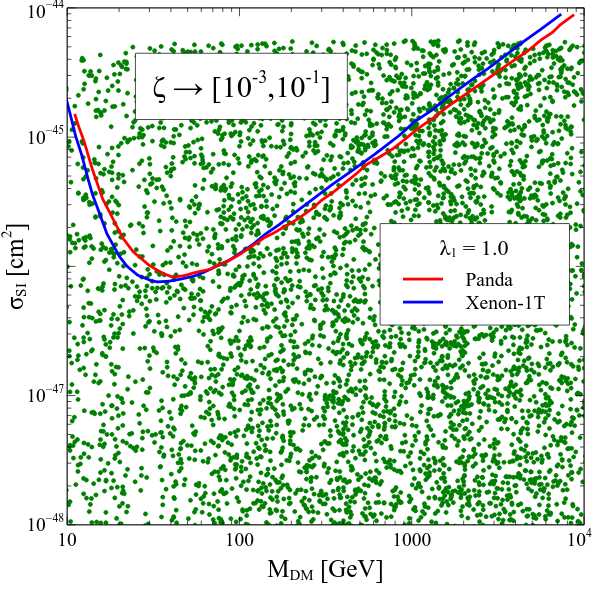}
\caption{Variation of $\sigma_{SI}$ with DM mass where the scalar
mixing angle $\zeta$ has been varied in the range as shown in legend.
Other parameters have been kept fixed $M_{h_1} = $ 125.5 GeV,
$M_{h_2} = 1.5$ TeV and $\lambda_1 = 1.0$.}
\label{mdm-sigma-si}
\end{figure}

The masses $M_{h_1}$, $M_{h_2}$ and mixing angle $\zeta$
are given in Eqs.\,(\ref{Eq:scalarmass}),
(\ref{Eq:mixing_angle}) while $\lambda_1$ is the Yukawa coupling
between ${\Psi_1}_L$, $\tilde{\phi_D}$ and ${\psi_1}_R$, related
as well to the Dark Matter mass by Eq.\, (\ref{dm-mass}).

In Fig.\,\ref{mdm-sigma-si}, we plot the variation
of the spin-independent DM-nucleon scattering cross section
with the DM mass. In generating the plot, we have kept $\lambda_1$ 
fixed at unity while DM mass has been varied in the range 10 GeV 
to $10^4$ GeV by appropriately adjusting $v_D$. 
The mixing angle $\zeta$ has been scanned over its present
allowed range i.e $\zeta\leq10^{-1}$ rad. From the
Fig.\,\,\ref{dd-fig}, we see that a part of the parameter space,
corresponding to large $ \zeta $ and heavy $ M_{h_2} $, is already 
ruled out by the null results at various direct detection experiments
like PandaX-II \cite{Cui:2017nnn} and Xenon1T \cite{ Aprile:2017iyp}, 
however enough parameter space is still open and can be tested in 
the future experiments like Darwin \cite{Aalbers:2016jon}.

\section{conclusion}
\label{sec:conclusion}

In this work we have tried to solve three major puzzles of cosmology by the presence of two RH neutrinos
and a Dark Sector charged under an $SU(2)_D$. The hidden sector of the model is chosen to 
resemble the SM electroweak sector, but with just two non mixing families, so that the mass of the 
DM particles could be similar to the SM fermions and the presence of the $SU(2)_D$ interaction is
crucial for annihilating away all the symmetric Dark Matter components.

We generate the neutrino mass through the Type-I seesaw mechanism, while both the lepton
(later processed into baryons) and a DM asymmetries are produced
from the decay of the lightest RH neutrino. Hence, all the three BSM
phenomena have a common origin. Moreover, the CP violation in both sectors is related to 
complex entries in the neutrino Yukawa couplings.
Two RH neutrinos with hierarchical masses are sufficient to accommodate all the 
present oscillation data and satisfy successfully the present day bounds on the sum
of the light neutrino masses. We have shown that the limited number of parameters
in the model results in strong correlations among some of the entries
in the Dirac and Majorana mass matrices.

Due to the strong hierarchy, we can generate the baryon asymmetry of the Universe and
an asymmetric Dark Matter component at the same time from the decay of the lighter RH 
neutrino state. Also in this case, two RH neutrino states are enough to give both, as 
long as they have similar Yukawa coupling with the light neutrinos and the additional 
dark fermions, so that the two decays can have naturally a similar decay rate and branching 
fraction.  As CP violation is sourced in both sectors by the neutrino Yukawa $y_{ij}$, 
 comparable values of $ \epsilon $ for leptons and the Dark Matter are expected in 
 most parameter space.
Indeed, the two CP asymmetries in the decays are exactly equal if the neutrino Dirac 
 mass has a purely imaginary column, which gives a real Majorana mass matrix for the
 light neutrinos and a vanishing Dirac phase. 
 We expect nevertheless to have one non-trivial Majorana phase, that could lead to a partial 
 cancellation in the matrix element for neutrinoless  $ \beta\beta $ decay. 
 Similarly the CP asymmetry are similar as well in the case $ \alpha_2 >> \alpha_1$
 for the Dark sector Yukawa with the RH neutrinos. 
 In other cases we have $ \epsilon_D < \epsilon_l $, but this can be partially compensated 
 by the presence of less effective wash-out processes or in any case by a 
 larger DM mass.
For specific choices of the parameters, i.e. the imaginary parts of the Yukawas 
and Dark Matter mass, we can simultaneously satisfy neutrino oscillation parameters 
bounds and produce the correct value of the baryon asymmetry and of the Dark Matter 
energy density. 

In this scenario the visible and dark sector do not communicate only through the neutrino
portal: indeed after the electroweak and dark $SU(2)_D $ symmetries are broken, a
mixing appears also in the scalar sector, so that the physical Higgs field contains also
a small component of Dark Higgs and can couple to Dark Matter.
For light Dark Matter we therefore expect an invisible contribution to the Higgs width
proportional to the mixing angle in the scalar sector.
If the Dark Matter is heavier than half of the Higgs mass, an observable signal could
still appear in the Direct Detection experiments and from the production of Dark Matter 
at colliders through an off-shell Higgs. Moreover, as in many models with an extended 
Higgs sector,  we can also expect to detect exotic scalars at colliders. 
Regarding the heavier Higgs state $ h_2$, Direct Detection and collider experiments are 
in our model highly complementary,  as for heavy $ M_{h_2} $, whereas the production of 
$h_2$ at colliders is suppressed,  the scattering cross section with nucleons becomes larger
for fixed mixing angle $\zeta$.
Note as well that in this type of model, as in all models with extended Higgs sector,
 the Higgs self-couplings  are modified in a characteristic way and that may be observable 
 even when all the Dark Sector particles are beyond the present collider reach,
 see e.g. \cite{Lopez-Val:2013yba, Ivanov:2017dad}.

\section{Acknowledgements} 
The authors would like to thank the Department of Atomic Energy
(DAE) Neutrino Project under the plan project of Harish-Chandra
Research Institute. AB would like to thank the organiser of
nu HoRIzons VII (http://www.hri.res.in/~nuhorizons/nuhri7/) for an invitation,
where this work was initiated. He also acknowledges
SERB, Govt.\,of\,\,India for financial support through NPDF fellowship
(PDF/2017/000490) during the early stage of this work. LC would
like to thank the Harish-Chandra Research Institute
for hospitality during the initial stages of this work.
This project has received funding from the European Union's Horizon
2020 research and innovation programme InvisiblesPlus RISE
under the Marie Sklodowska-Curie grant  agreement  No  690575.
This  project  has received  funding  from  the  European
Union's Horizon  2020  research  and  innovation programme
Elusives  ITN under the  Marie  Sklodowska-Curie grant
agreement No 674896.

\appendix
\section{Expression for the Majorana mass matrix of light neutrinos}
\label{app:mnu}
Here we have given the expression of all the elements of the light neutrino
mass matrix $m_{\nu}$ (using Eq.\,\ref{neutrino-mass}) in terms of the Yukawa
couplings and the RH neutrino masses.
\begin{eqnarray}
&& (\tilde m_{\nu})_{11} = \frac{y_{ee}^2}{M_{N_1}} +
\frac{(y_{e\mu}^{R})^2 - (y_{e\mu}^{I})^2}{M_{N_2}} -
i \frac{2\, y_{e\mu}^R y_{e\mu}^I}{M_{N_2}},
\nn \\
&& (\tilde m_{\nu})_{12} = \frac{y_{ee} y_{\mu e}}{M_{N_1}}
+ \frac{y_{e\mu}^R y_{\mu\mu}^R - y_{e\mu}^I y_{\mu\mu}^I}{M_{N_2}}
- i \frac{y_{e\mu}^R y_{\mu\mu}^I + y_{e\mu}^I y_{\mu\mu}^R}{M_{N_2}}, \nn \\
&& (\tilde m_{\nu})_{13} =\frac{y_{ee} y_{\tau e}}{M_{N_1}}
+ \frac{y_{e\mu}^R y_{\tau\mu}^R - y_{e\mu}^I y_{\tau\mu}^I}{M_{N_2}}
- i \frac{y_{e\mu}^I y_{\tau\mu}^R + y_{e\mu}^R y_{\tau\mu}^I}{M_{N_2}}, \nn \\
&& (\tilde m_{\nu})_{21} = (\tilde m_{\nu})_{12}, \nn \\
&& (\tilde m_{\nu})_{22} = \frac{y_{\mu e}^2}{M_{N_1}} +
\frac{(y_{\mu\mu}^{R})^2 - (y_{\mu\mu}^{I})^2}{M_{N_2}} -
i \frac{2\, y_{\mu\mu}^R y_{\mu\mu}^I}{M_{N_2}} \nn \\
&& (\tilde m_{\nu})_{23} = \frac{y_{\mu e} y_{\tau e}}{M_{N_1}}
+ \frac{y_{\mu\mu}^R y_{\tau\mu}^R - y_{\mu\mu}^I y_{\tau\mu}^I}{M_{N_2}}
- i \frac{y_{\mu\mu}^I y_{\tau\mu}^R + y_{\mu\mu}^R y_{\tau\mu}^I}{M_{N_2}}, \nn \\
&& (\tilde m_{\nu})_{31} =  (\tilde m_{\nu})_{13}, \nn \\
&& (\tilde m_{\nu})_{32} = (\tilde m_{\nu})_{23}, \nn \\
&&(\tilde m_{\nu})_{33} = \frac{y_{\tau e}^2}{M_{N_1}} +
\frac{(y_{\tau\mu}^{R})^2 - (y_{\tau\mu}^{I})^2}{M_{N_2}} -
i \frac{2\, y_{\tau\mu}^R y_{\tau\mu}^I}{M_{N_2}}, 
 \label{mnuelements} \\ 
 &&m_{\nu} = -\frac{v^2}{2}
 \left(\begin{array}{ccc}
(\tilde m_{\nu})_{11} ~~&~~ (\tilde m_{\nu})_{12}
~~&~~(\tilde m_{\nu})_{13}\\
(\tilde m_{\nu})_{21}
~~&~~ (\tilde m_{\nu})_{22}
~~&~~(\tilde m_{\nu})_{23}\\
(\tilde m_{\nu})_{31} ~~&
~~ (\tilde m_{\nu})_{32} ~~&~~ (\tilde m_{\nu})_{33}\\
\end{array}\right) \,.
\label{neutrino-mass-elements}
\end{eqnarray}

We see from these expressions that if the second column of the Yukawa matrix is imaginary,
i.e. for $ y_{i \mu}^{R} = 0 $, the light neutrino mass is the sum of two degenerated real matrices, 
each with a single non-zero eigenvalue and opposite sign. 
If the massive eigenvectors of the two matrices are orthogonal to each other, 
we have then simply two opposite-sign mass  eigenstates and one zero mass eigenstate as:
\begin{eqnarray}
m_3 &=& - \frac{v^2}{2} \frac{\sum_i y_{i e}^2 }{M_1} \; ,
\\
m_2 &=& \frac{v^2}{2} \frac{\sum_i (y_{i \mu}^{I})^2 }{M_2} \; ,
\\
m_1 &=& 0\; .
\end{eqnarray}
So we have to choose the hierarchy in $ M_i $ and the Yukawa couplings
appropriately in order to match the measured mass differences.
If the two massive eigenvectors are not orthogonal, a more complex mixing pattern appears
and the two mass eigenstates obtain contributions from both heavy RH neutrinos, nevertheless
for hierarchical masses and not so strongly hierarchical Yukawas, still the heaviest mass 
$ m_3 $ is mostly determined by the lighter mass $ M_1 $.
Generically for real $ m_{\nu} $, the mixing matrix is real, so that the Dirac phase is exactly 
vanishing, but the Majorana phases are not, as they have to be chosen to give positive 
light neutrino masses, i.e. we obtain for the mass eigenstates above the Majorana phases 
$ \alpha_{31} = \pi, \alpha_{21} = 0 $. 
We expect also in the generic case with a purely imaginary Yukawa column to have two 
eigenstates with different Majorana phases, leading to a partial cancellation in the matrix 
element for neutrinoless double beta decay:
\begin{equation}
m_{\beta\beta} = | \sum_i m_i U_{ei}^2 | = | m_3 \sin^2\theta_{13} - m_2 \cos^2 \theta_{13} \sin^2 \theta_{12} |
\sim 10^{-2} \mbox{eV} \; .
\end{equation}

\section{Expression of $\gamma_{D_1}$, $\gamma_{\phi_h,s}^1$
and $\gamma^1_{\phi_h,t}$:}
\label{app:b}
Expression of $\gamma_{D_1}$ takes the following form \cite{Plumacher:1996kc},
\begin{eqnarray}
\gamma_{D_1} = n_{N_1}^{eq} \frac{K_{1}(Z)}{K_{2}(z)} \Gamma_{N_1}\,\,,
\end{eqnarray}
where $n_{N_1}^{eq}$ is the equilibrium number density of the RH neutrino $N_1$
and $K_n(z)$ is the $n$th order modified Bessel function of second kind
while $\Gamma_{N_1}$ is the total decay width of $N_1$. The expression
of $\Gamma_{N_1}$ is given in Eq.\,\,(\ref{Eq:gamma_tot_N1}). 

Further, the general expression of $\gamma(a+b \leftrightarrow i + j + ...)$
for a two body scattering process $a+b \leftrightarrow i + j + ...$
is given by \cite{Plumacher:1996kc},
\begin{eqnarray}
\gamma(a+b \leftrightarrow i + j + ...)  = \frac{T}{64 \pi^4}
\int_{(M_{a} + M_{b})^2}^{\infty} ds\, \hat{\sigma}(s) \sqrt{s}\,
K_1\left(\frac{\sqrt{s}}{T}\right)\,\,,
\end{eqnarray}
where, $s$ is one of the Mandelstam variables which physically represents
square of the centre of mass energy for a scattering process in
centre of momentum frame. Moreover, $\hat{\sigma}(s)$ is the reduced
cross section for the scattering process $a+b \leftrightarrow i + j + ...$,
which is related to the actual cross section by the following relation
\begin{eqnarray}
\hat{\sigma}(s) = \dfrac{8}{s} \left[\left(p_{a}.p_{b}\right)^2 -
M^2_a M^2_b\right]\sigma(s)\,.
\end{eqnarray}
Here, $p_i$ and $M_i$ are the three momentum and mass of the species $i$
respectively. The expressions of reduced cross sections for the processes
$N_1 + l \rightarrow \bar{t} + q$ ($s$-channel process mediated by $\phi_h$)
and $N_1 + t \rightarrow \bar{l} + q$ ($t$-channel process mediated by $\phi_h$)
are given as 
\begin{eqnarray}
\hat{\sigma}_{\phi_h,s} &=& \frac{3 \pi \alpha^2 M_{t}^2}{M_{W}^{4} \sin^4\theta_{w}}
(M_{D}^{\dagger} M_{D})_{11} \left[\frac{s - M_{N_1}^2}{s} \right]^2\,, \nn \\
\hat{\sigma}_{\phi_h,t} &=& \frac{3 \pi \alpha^2 M_{t}^2}{M_{W}^{4} \sin^4\theta_{w}}
(M_{D}^{\dagger} M_{D})_{11} \left[\frac{s - M_{N_1}^2}{s} + \frac{M_{N_1}^2}{s}
\ln\left(\frac{s- M_{N_1}^2 + M_{h_1}^2}{M_{h_1}^2}\right) \right]\,\,, 
\end{eqnarray}
where $\alpha = \dfrac{g^2 \sin^{2} \theta_{w}}{4\pi}$, $g$ being the SU(2)$_{L}$ gauge coupling
and $\theta_w$ is the weak mixing angle (Weinberg angle).


\begin{thebibliography}{99}


\bibitem{Cowan:1992xc} 
C.~L.~Cowan, F.~Reines, F.~B.~Harrison, H.~W.~Kruse and A.~D.~McGuire,
``{\it Detection of the free neutrino: A Confirmation}'',
Science {\bf 124}, 103 (1956).
\bibitem{Fukuda:1998mi} 
Y.~Fukuda {\it et al.} [Super-Kamiokande Collaboration],
``{\it Evidence for oscillation of atmospheric neutrinos}'',
Phys.\ Rev.\ Lett.\  {\bf 81}, 1562 (1998)
[hep-ex/9807003].
\bibitem{Ahmad:2002jz} 
Q.~R.~Ahmad {\it et al.} [SNO Collaboration],
``{\it Direct evidence for neutrino flavor transformation
from neutral current interactions in the Sudbury Neutrino Observatory}'',
Phys.\ Rev.\ Lett.\  {\bf 89}, 011301 (2002)
[nucl-ex/0204008].
\bibitem{Eguchi:2002dm} 
K.~Eguchi {\it et al.} [KamLAND Collaboration],
``{\it First results from KamLAND: Evidence for
reactor anti-neutrino disappearance}'',
Phys.\ Rev.\ Lett.\  {\bf 90}, 021802 (2003)
[hep-ex/0212021].
\bibitem{An:2015nua} 
  F.~P.~An {\it et al.} [Daya Bay Collaboration],
  ``{\it Measurement of the Reactor Antineutrino Flux and Spectrum at Daya Bay}'',
  Phys.\ Rev.\ Lett.\  {\bf 116}, no. 6, 061801 (2016)
  [arXiv:1508.04233 [hep-ex]].

\bibitem{RENO:2015ksa} 
  J.~H.~Choi {\it et al.} [RENO Collaboration],
  ``{\it Observation of Energy and Baseline Dependent Reactor
  Antineutrino Disappearance in the RENO Experiment}'',
  Phys.\ Rev.\ Lett.\  {\bf 116}, no. 21, 211801 (2016)
  [arXiv:1511.05849 [hep-ex]].
\bibitem{Abe:2014bwa} 
  Y.~Abe {\it et al.} [Double Chooz Collaboration],
  ``{\it Improved measurements of the neutrino mixing angle
  $\theta_{13}$ with the Double Chooz detector}'',
  JHEP {\bf 1410}, 086 (2014)
  Erratum: [JHEP {\bf 1502}, 074 (2015)]
  [arXiv:1406.7763 [hep-ex]].
\bibitem{Abe:2015awa} 
  K.~Abe {\it et al.} [T2K Collaboration],
  ``{\it Measurements of neutrino oscillation in appearance
  and disappearance channels by the T2K experiment with 6.6$\times$10$^{20}$ protons on target}'',
  Phys.\ Rev.\ D {\bf 91}, no. 7, 072010 (2015)
  [arXiv:1502.01550 [hep-ex]].

\bibitem{Salzgeber:2015gua} 
  M.~Ravonel Salzgeber [T2K Collaboration],
  ``{\it Anti-neutrino oscillations with T2K}'',
  arXiv:1508.06153 [hep-ex].
\bibitem{Adamson:2016tbq} 
  P.~Adamson {\it et al.} [NOvA Collaboration],
  ``{\it First measurement of electron neutrino appearance in NOvA}'',
  Phys.\ Rev.\ Lett.\  {\bf 116}, no. 15, 151806 (2016)
  [arXiv:1601.05022 [hep-ex]].
\bibitem{Adamson:2016xxw} 
  P.~Adamson {\it et al.} [NOvA Collaboration],
  ``{\it First measurement of muon-neutrino disappearance in NOvA}'',
  Phys.\ Rev.\ D {\bf 93}, no. 5, 051104 (2016)
  [arXiv:1601.05037 [hep-ex]].

\bibitem{Sofue:2000jx} 
Y.~Sofue and V.~Rubin,
``{\it Rotation curves of spiral galaxies}'',
Ann.\ Rev.\ Astron.\ Astrophys.\  {\bf 39}, 137 (2001)
[astro-ph/0010594].
\bibitem{Clowe:2003tk} 
D.~Clowe, A.~Gonzalez and M.~Markevitch,
``{\it Weak lensing mass reconstruction of the interacting
cluster 1E0657-558: Direct evidence for the existence of dark matter}'',
Astrophys.\ J.\  {\bf 604}, 596 (2004)
[astro-ph/0312273].
\bibitem{Harvey:2015hha} 
D.~Harvey, R.~Massey, T.~Kitching, A.~Taylor and E.~Tittley,
``{\it The non-gravitational interactions of dark matter
in colliding galaxy clusters}'',
Science {\bf 347}, 1462 (2015)
[arXiv:1503.07675 [astro-ph.CO]].

\bibitem{Bartelmann:1999yn} 
M.~Bartelmann and P.~Schneider,
``{\it Weak gravitational lensing}'',
Phys.\ Rept.\  {\bf 340}, 291 (2001)
[astro-ph/9912508].


\bibitem{Hinshaw:2012aka} 
G.~Hinshaw {\it et al.} [WMAP Collaboration],
``{\it Nine-Year Wilkinson Microwave Anisotropy Probe (WMAP)
Observations: Cosmological Parameter Results}'',
Astrophys.\ J.\ Suppl.\  {\bf 208}, 19 (2013)
[arXiv:1212.5226 [astro-ph.CO]].

\bibitem{Ade:2015xua}
  P.~A.~R.~Ade {\it et al.} [Planck Collaboration],
  ``{\it Planck 2015 results. XIII. Cosmological parameters}",
  Astron.\ Astrophys.\  {\bf 594} (2016) A13
  doi:10.1051/0004-6361/201525830
  [arXiv:1502.01589 [astro-ph.CO]].

\bibitem{Aghanim:2018eyx}
  N.~Aghanim {\it et al.} [Planck Collaboration],
 " {\it Planck 2018 results. VI. Cosmological parameters}",
  arXiv:1807.06209 [astro-ph.CO].

\bibitem{Taoso:2007qk} 
  M.~Taoso, G.~Bertone and A.~Masiero,
  ``{\it Dark Matter Candidates: A Ten-Point Test}'',
  JCAP {\bf 0803}, 022 (2008)
  [arXiv:0711.4996 [astro-ph]].

\bibitem{Choudhury:2018byy}
  S.~Roy Choudhury and S.~Choubey,
  JCAP {\bf 1809} (2018) no.09,  017
  doi:10.1088/1475-7516/2018/09/017
  [arXiv:1806.10832 [astro-ph.CO]].

\bibitem{Sakharov:1967dj} 
  A.~D.~Sakharov,
  ``{\it Violation of CP Invariance, C asymmetry, and baryon asymmetry of the universe}'',
  Pisma Zh.\ Eksp.\ Teor.\ Fiz.\  {\bf 5}, 32 (1967)
  [JETP Lett.\  {\bf 5}, 24 (1967)]
  [Sov.\ Phys.\ Usp.\  {\bf 34}, no. 5, 392 (1991)]
  [Usp.\ Fiz.\ Nauk {\bf 161}, no. 5, 61 (1991)].

\bibitem{Schechter:1980gr} 
  J.~Schechter and J.~W.~F.~Valle,
  ``{\it Neutrino Masses in SU(2) x U(1) Theories}'',
  Phys.\ Rev.\ D {\bf 22}, 2227 (1980).
\bibitem{Mohapatra:1979ia} 
  R.~N.~Mohapatra and G.~Senjanovic,
  ``{\it Neutrino Mass and Spontaneous Parity Violation}'',
  Phys.\ Rev.\ Lett.\  {\bf 44}, 912 (1980).

\bibitem{Fukugita:1986hr}
  M.~Fukugita and T.~Yanagida,
  Phys.\ Lett.\ B {\bf 174} (1986) 45.
  doi:10.1016/0370-2693(86)91126-3

\bibitem{Covi:1996wh}
  L.~Covi, E.~Roulet and F.~Vissani,
  Phys.\ Lett.\ B {\bf 384} (1996) 169
  doi:10.1016/0370-2693(96)00817-9
  [hep-ph/9605319].

\bibitem{Kuzmin:1985mm}
  V.~A.~Kuzmin, V.~A.~Rubakov and M.~E.~Shaposhnikov,
  Phys.\ Lett.\  {\bf 155B} (1985) 36.
  doi:10.1016/0370-2693(85)91028-7

\bibitem{An:2009vq} 
  H.~An, S.~L.~Chen, R.~N.~Mohapatra and Y.~Zhang,
  ``{\it Leptogenesis as a Common Origin for Matter and Dark Matter}'',
  JHEP {\bf 1003}, 124 (2010)
  [arXiv:0911.4463 [hep-ph]].

\bibitem{Dutta:2010va} 
  B.~Dutta and J.~Kumar,
  ``{\it Asymmetric Dark Matter from Hidden Sector Baryogenesis}'',
  Phys.\ Lett.\ B {\bf 699}, 364 (2011)
  [arXiv:1012.1341 [hep-ph]].


\bibitem{Falkowski:2011xh} 
  A.~Falkowski, J.~T.~Ruderman and T.~Volansky,
  ``{\it Asymmetric Dark Matter from Leptogenesis}'',
  JHEP {\bf 1105}, 106 (2011)
  [arXiv:1101.4936 [hep-ph]].

\bibitem{Graesser:2011wi} 
  M.~L.~Graesser, I.~M.~Shoemaker and L.~Vecchi,
  ``{\it Asymmetric WIMP dark matter}'',
  JHEP {\bf 1110}, 110 (2011)
  [arXiv:1103.2771 [hep-ph]].

\bibitem{McDermott:2011jp} 
  S.~D.~McDermott, H.~B.~Yu and K.~M.~Zurek,
  ``{\it Constraints on Scalar Asymmetric Dark Matter from Black Hole Formation in Neutron Stars}'',
  Phys.\ Rev.\ D {\bf 85}, 023519 (2012)
  [arXiv:1103.5472 [hep-ph]].

\bibitem{Iminniyaz:2011yp} 
  H.~Iminniyaz, M.~Drees and X.~Chen,
  ``{\it Relic Abundance of Asymmetric Dark Matter}'',
  JCAP {\bf 1107}, 003 (2011)
  [arXiv:1104.5548 [hep-ph]].

\bibitem{Kouvaris:2011fi} 
  C.~Kouvaris and P.~Tinyakov,
  ``{\it Excluding Light Asymmetric Bosonic Dark Matter}'',
  Phys.\ Rev.\ Lett.\  {\bf 107}, 091301 (2011)
  [arXiv:1104.0382 [astro-ph.CO]].

\bibitem{Arina:2011cu} 
  C.~Arina and N.~Sahu,
  ``{\it Asymmetric Inelastic Inert Doublet Dark Matter from Triplet Scalar Leptogenesis}'',
  Nucl.\ Phys.\ B {\bf 854}, 666 (2012)
  [arXiv:1108.3967 [hep-ph]].

\bibitem{Buckley:2011ye} 
  M.~R.~Buckley and S.~Profumo,
  ``{\it Regenerating a Symmetry in Asymmetric Dark Matter}'',
  Phys.\ Rev.\ Lett.\  {\bf 108}, 011301 (2012)
  [arXiv:1109.2164 [hep-ph]].

\bibitem{Lin:2011gj} 
  T.~Lin, H.~B.~Yu and K.~M.~Zurek,
  ``{\it On Symmetric and Asymmetric Light Dark Matter}'',
  Phys.\ Rev.\ D {\bf 85}, 063503 (2012)
  [arXiv:1111.0293 [hep-ph]].
  
\bibitem{Blum:2012nf} 
  K.~Blum, A.~Efrati, Y.~Grossman, Y.~Nir and A.~Riotto,
  ``{\it Asymmetric Higgsino Dark Matter}'',
  Phys.\ Rev.\ Lett.\  {\bf 109}, 051302 (2012)
  [arXiv:1201.2699 [hep-ph]].
  
\bibitem{Blennow:2012de} 
  M.~Blennow, E.~Fernandez-Martinez, O.~Mena, J.~Redondo and P.~Serra,
  ``{\it Asymmetric Dark Matter and Dark Radiation}'',
  JCAP {\bf 1207}, 022 (2012)
  [arXiv:1203.5803 [hep-ph]].
  
\bibitem{Okada:2012rm} 
  N.~Okada and O.~Seto,
  ``{\it Originally Asymmetric Dark Matter}'',
  Phys.\ Rev.\ D {\bf 86}, 063525 (2012)
  [arXiv:1205.2844 [hep-ph]].
    
\bibitem{Perez:2013nra} 
  P.~Fileviez Perez and M.~B.~Wise,
  ``{\it Baryon Asymmetry and Dark Matter Through the Vector-Like Portal}'',
  JHEP {\bf 1305}, 094 (2013)
  [arXiv:1303.1452 [hep-ph]].

\bibitem{Petraki:2013wwa} 
  K.~Petraki and R.~R.~Volkas,
  ``{\it Review of asymmetric dark matter}'',
  Int.\ J.\ Mod.\ Phys.\ A {\bf 28}, 1330028 (2013)
  [arXiv:1305.4939 [hep-ph]].

\bibitem{Bhattacherjee:2013jca} 
  B.~Bhattacherjee, S.~Matsumoto, S.~Mukhopadhyay and M.~M.~Nojiri,
  ``{\it Phenomenology of light fermionic asymmetric dark matter}'',
  JHEP {\bf 1310}, 032 (2013)
  [arXiv:1306.5878 [hep-ph]].

\bibitem{Zurek:2013wia} 
  K.~M.~Zurek,
  ``{\it Asymmetric Dark Matter: Theories, Signatures, and Constraints}'',
  Phys.\ Rept.\  {\bf 537}, 91 (2014)
  [arXiv:1308.0338 [hep-ph]].


\bibitem{Zhao:2014nsa} 
  Y.~Zhao and K.~M.~Zurek,
  ``{\it Indirect Detection Signatures for the Origin of Asymmetric Dark Matter}'',
  JHEP {\bf 1407}, 017 (2014)
  [arXiv:1401.7664 [hep-ph]].
  
\bibitem{Bishara:2014gwa} 
  F.~Bishara and J.~Zupan,
  ``{\it Continuous Flavor Symmetries and the Stability of Asymmetric Dark Matter}'',
  JHEP {\bf 1501}, 089 (2015)
  [arXiv:1408.3852 [hep-ph]].
  
\bibitem{Hamze:2014wca} 
  A.~Hamze, C.~Kilic, J.~Koeller, C.~Trendafilova and J.~H.~Yu,
  ``{\it Lepton-Flavored Asymmetric Dark Matter and Interference in Direct Detection}'',
  Phys.\ Rev.\ D {\bf 91}, no. 3, 035009 (2015)
  [arXiv:1410.3030 [hep-ph]].

\bibitem{Ibarra:2016fco} 
  A.~Ibarra, S.~Lopez-Gehler, E.~Molinaro and M.~Pato,
  ``{\it Gamma-ray triangles: a possible signature of asymmetric dark matter in indirect searches}'',
  Phys.\ Rev.\ D {\bf 94}, no. 10, 103003 (2016)
  [arXiv:1604.01899 [hep-ph]].

\bibitem{Narendra:2018vfw} 
  N.~Narendra, S.~Patra, N.~Sahu and S.~Shil,
  ``{\it Baryogenesis via Leptogenesis from Asymmetric Dark Matter and radiatively generated Neutrino mass}'',
  arXiv:1805.04860 [hep-ph].

\bibitem{Dong:2018aak} 
  P.~V.~Dong, D.~T.~Huong, D.~A.~Camargo, F.~S.~Queiroz and J.~W.~F.~Valle,
  ``{\it Asymmetric Dark Matter, Inflation and Leptogenesis from B-L Symmetry Breaking}'',
  arXiv:1805.08251 [hep-ph].

\bibitem{Ibe:2018juk} 
  M.~Ibe, A.~Kamada, S.~Kobayashi and W.~Nakano,
  JHEP {\bf 1811}, 203 (2018)
  [arXiv:1805.06876 [hep-ph]].

\bibitem{Dessert:2018khu} 
  C.~Dessert, C.~Kilic, C.~Trendafilova and Y.~Tsai,
  arXiv:1811.05534 [hep-ph].

\bibitem{Ibe:2018tex} 
  M.~Ibe, A.~Kamada, S.~Kobayashi, T.~Kuwahara and W.~Nakano,
  arXiv:1811.10232 [hep-ph].


\bibitem{Banik:2015aya} 
  A.~Dutta Banik, D.~Majumdar and A.~Biswas,
  Eur.\ Phys.\ J.\ C {\bf 76}, no. 6, 346 (2016)
  [arXiv:1506.05665 [hep-ph]].

\bibitem{Witten:1982fp} 
  E.~Witten,
  ``{\it An SU(2) Anomaly}'',
  Phys.\ Lett.\ B {\bf 117}, 324 (1982)
  [Phys.\ Lett.\  {\bf 117B}, 324 (1982)].



\bibitem{Akerib:2015rjg} 
D.~S.~Akerib {\it et al.} [LUX Collaboration],
``{\it Improved Limits on Scattering of Weakly Interacting
Massive Particles from Reanalysis of 2013 LUX Data}'',
Phys.\ Rev.\ Lett.\  {\bf 116}, no. 16, 161301 (2016)
[arXiv:1512.03506 [astro-ph.CO]].
\bibitem{Aprile:2015uzo} 
E.~Aprile {\it et al.} [XENON Collaboration],
``{\it Physics reach of the XENON1T dark matter experiment}'',
JCAP {\bf 1604}, no. 04, 027 (2016)
[arXiv:1512.07501 [physics.ins-det]].
\bibitem{Agnese:2014aze} 
R.~Agnese {\it et al.} [SuperCDMS Collaboration],
``{\it Search for Low-Mass Weakly Interacting Massive
Particles with SuperCDMS}'',
Phys.\ Rev.\ Lett.\  {\bf 112}, no. 24, 241302 (2014)
[arXiv:1402.7137 [hep-ex]].
\bibitem{lux2016}
Talk by A.~Manalaysay for the LUX collaboration,\\
https://idm2016.shef.ac.uk/indico/event/0/contribution/50/material/slides/0.pdf

\bibitem{Aalbers:2016jon} 
J.~Aalbers {\it et al.} [DARWIN Collaboration],
``{\it DARWIN: towards the ultimate dark matter detector}'',
arXiv:1606.07001 [astro-ph.IM].

\bibitem{Aprile:2017iyp} 
  E.~Aprile {\it et al.} [XENON Collaboration],
  ``{\it First Dark Matter Search Results from the XENON1T Experiment}'',
  Phys.\ Rev.\ Lett.\  {\bf 119}, no. 18, 181301 (2017)
  [arXiv:1705.06655 [astro-ph.CO]].

\bibitem{Cui:2017nnn} 
  X.~Cui {\it et al.} [PandaX-II Collaboration],
  ``{\it Dark Matter Results From 54-Ton-Day Exposure of PandaX-II Experiment}'',
  Phys.\ Rev.\ Lett.\  {\bf 119}, no. 18, 181302 (2017)
  [arXiv:1708.06917 [astro-ph.CO]].




\bibitem{Falkowski:2009yz}
  A.~Falkowski, J.~Juknevich and J.~Shelton,
  arXiv:0908.1790 [hep-ph].
  
\bibitem{Gonzalez-Macias:2016vxy}
  V.~Gonzalez-Macias, J.~I.~Illana and J.~Wudka,
  JHEP {\bf 1605} (2016) 171
  doi:10.1007/JHEP05(2016)171
  [arXiv:1601.05051 [hep-ph]].
  
\bibitem{Escudero:2016tzx}
  M.~Escudero, N.~Rius and V.~Sanz,
  JHEP {\bf 1702} (2017) 045
  doi:10.1007/JHEP02(2017)045
  [arXiv:1606.01258 [hep-ph]].
  
\bibitem{Escudero:2016ksa}
  M.~Escudero, N.~Rius and V.~Sanz,
  Eur.\ Phys.\ J.\ C {\bf 77} (2017) no.6,  397
  doi:10.1140/epjc/s10052-017-4963-x
  [arXiv:1607.02373 [hep-ph]].
  
\bibitem{Batell:2017cmf}
  B.~Batell, T.~Han, D.~McKeen and B.~Shams Es Haghi,
  Phys.\ Rev.\ D {\bf 97} (2018) no.7,  075016
  doi:10.1103/PhysRevD.97.075016
  [arXiv:1709.07001 [hep-ph]].
  
\bibitem{Chianese:2018dsz}
  M.~Chianese and S.~F.~King,
  JCAP {\bf 1809} (2018) no.09,  027
  doi:10.1088/1475-7516/2018/09/027
  [arXiv:1806.10606 [hep-ph]].

\bibitem{Khachatryan:2016vau} 
  G.~Aad {\it et al.} [ATLAS and CMS Collaborations],
  JHEP {\bf 1608}, 045 (2016)
  doi:10.1007/JHEP08(2016)045
  [arXiv:1606.02266 [hep-ex]].

\bibitem{ATLAS:2017ovn}
  The ATLAS collaboration [ATLAS Collaboration],
  ATLAS-CONF-2017-047.

\bibitem{CMS:2018lkl}
  CMS Collaboration [CMS Collaboration],
  CMS-PAS-HIG-17-031.


\bibitem{Capozzi:2016rtj} 
  F.~Capozzi, E.~Lisi, A.~Marrone, D.~Montanino and A.~Palazzo,
  ``{\it Neutrino masses and mixings: Status of known and unknown $3\nu$ parameters}'',
  Nucl.\ Phys.\ B {\bf 908}, 218 (2016)
  [arXiv:1601.07777 [hep-ph]].

\bibitem{Abe:2011sj} 
  K.~Abe {\it et al.} [T2K Collaboration],
  ``{\it Indication of Electron Neutrino Appearance from an Accelerator-produced Off-axis Muon Neutrino Beam}'',
  Phys.\ Rev.\ Lett.\  {\bf 107}, 041801 (2011)
  [arXiv:1106.2822 [hep-ex]].

\bibitem{Abe:2015zbg} 
  K.~Abe {\it et al.} [Hyper-Kamiokande Proto- Collaboration],
  ``{\it Physics potential of a long-baseline neutrino oscillation experiment using a J-PARC neutrino beam and Hyper-Kamiokande}'',
  PTEP {\bf 2015}, 053C02 (2015)
  [arXiv:1502.05199 [hep-ex]].


\bibitem{Acciarri:2015uup} 
  R.~Acciarri {\it et al.} [DUNE Collaboration],
  ``{\it Long-Baseline Neutrino Facility (LBNF) and Deep Underground Neutrino Experiment (DUNE) : Conceptual Design Report, Volume 2: The Physics Program for DUNE at LBNF}'',
  arXiv:1512.06148 [physics.ins-det].
  
\bibitem{Acciarri:2016ooe} 
  R.~Acciarri {\it et al.} [DUNE Collaboration],
  ``{\it Long-Baseline Neutrino Facility (LBNF) and Deep Underground Neutrino Experiment (DUNE) : Conceptual Design Report, Volume 4 The DUNE Detectors at LBNF}'',
  arXiv:1601.02984 [physics.ins-det].
  
\bibitem{Strait:2016mof} 
  J.~Strait {\it et al.} [DUNE Collaboration],
  ``{\it Long-Baseline Neutrino Facility (LBNF) and Deep Underground Neutrino Experiment (DUNE) : Conceptual Design Report, Volume 3: Long-Baseline Neutrino Facility
  for DUNE June 24, 2015}'',
  arXiv:1601.05823 [physics.ins-det].

\bibitem{Acciarri:2016crz} 
  R.~Acciarri {\it et al.} [DUNE Collaboration],
  ``{\it Long-Baseline Neutrino Facility (LBNF) and Deep Underground Neutrino Experiment (DUNE) : Conceptual Design Report, Volume 1: The LBNF and DUNE Projects}'',
  arXiv:1601.05471 [physics.ins-det].

\bibitem{Kumar:2017sdq} 
  S.~Ahmed {\it et al.} [ICAL Collaboration],
  ``{\it Physics Potential of the ICAL detector at the India-based Neutrino Observatory (INO)}'',
  Pramana {\bf 88}, no. 5, 79 (2017)
  [arXiv:1505.07380 [physics.ins-det]].


\bibitem{Abe:2018wpn} 
  K.~Abe {\it et al.} [T2K Collaboration],
  Phys.\ Rev.\ Lett.\  {\bf 121}, no. 17, 171802 (2018)
  doi:10.1103/PhysRevLett.121.171802
  [arXiv:1807.07891 [hep-ex]].

\bibitem{NOvA:2018gge} 
  M.~A.~Acero {\it et al.} [NOvA Collaboration],
  Phys.\ Rev.\ D {\bf 98}, 032012 (2018)
  doi:10.1103/PhysRevD.98.032012
  [arXiv:1806.00096 [hep-ex]].

\bibitem{Plumacher:1996kc} 
  M.~Plumacher,
  ``{\it Baryogenesis and lepton number violation}'',
  Z.\ Phys.\ C {\bf 74}, 549 (1997)
  [hep-ph/9604229].

\bibitem{Iso:2010mv} 
  S.~Iso, N.~Okada and Y.~Orikasa,
  ``{\it Resonant Leptogenesis in the Minimal B-L Extended Standard Model at TeV}'',
  Phys.\ Rev.\ D {\bf 83}, 093011 (2011)
  [arXiv:1011.4769 [hep-ph]].

\bibitem{Khlebnikov:1988sr} 
  S.~Y.~Khlebnikov and M.~E.~Shaposhnikov,
  ``{\it The Statistical Theory of Anomalous Fermion Number Nonconservation}'',
  Nucl.\ Phys.\ B {\bf 308}, 885 (1988).

\bibitem{Edsjo:1997bg} 
  J.~Edsjo and P.~Gondolo,
  ``{\it Neutralino relic density including coannihilations}'',
  Phys.\ Rev.\ D {\bf 56}, 1879 (1997)
  [hep-ph/9704361].

\bibitem{ArkaniHamed:2006mb}
  N.~Arkani-Hamed, A.~Delgado and G.~F.~Giudice,
  ``{\it The Well-tempered neutralino}",
  Nucl.\ Phys.\ B {\bf 741} (2006) 108
  doi:10.1016/j.nuclphysb.2006.02.010
  [hep-ph/0601041].

\bibitem{Baldes:2017gzw}
  I.~Baldes and K.~Petraki,
  JCAP {\bf 1709} (2017) no.09,  028
  doi:10.1088/1475-7516/2017/09/028
  [arXiv:1703.00478 [hep-ph]].


\bibitem{DEramo:2010keq}
  F.~D'Eramo and J.~Thaler,
 "{\it Semi-annihilation of Dark Matter}'',
  JHEP {\bf 1006} (2010) 109
  [arXiv:1003.5912 [hep-ph]].

\bibitem{Cline:2013gha} 
 J.~M.~Cline, K.~Kainulainen, P.~Scott and C.~Weniger,
``{\it Update on scalar singlet dark matter}'',
Phys.\ Rev.\ D {\bf 88}, 055025 (2013)
Erratum: [Phys.\ Rev.\ D {\bf 92}, no. 3, 039906 (2015)]
[arXiv:1306.4710 [hep-ph]].

\bibitem{Lopez-Val:2013yba}
  D.~Lopez-Val, T.~Plehn and M.~Rauch,
  "{\it Measuring extended Higgs sectors as a consistent free couplings model}'',
  JHEP {\bf 1310} (2013) 134
  [arXiv:1308.1979 [hep-ph]].

\bibitem{Ivanov:2017dad}
  I.~P.~Ivanov,
  "{\it Building and testing models with extended Higgs sectors}",
  Prog.\ Part.\ Nucl.\ Phys.\  {\bf 95} (2017) 160
  [arXiv:1702.03776 [hep-ph]].


\end{thebibliography}
\end{document}